\documentclass[pdflatex,11pt,twocolumn,tighten,times]{aastex631}
\usepackage{bm}
\usepackage{color}
\usepackage{graphicx}

\accepted{2021 October 29}

\submitjournal{ApJ}
\shorttitle{Convection and Dynamo in Newly-born Neutron Stars}
\shortauthors{Masada et al.}
\begin{document}
\title{Convection and Dynamo in Newly-born Neutron Stars}
\correspondingauthor{Youhei MASADA}
\email{ymasada@auecc.aichi-edu.ac.jp}

\author[0000-0002-0786-7307]{Youhei Masada}
\affiliation{Department of Science Education, Aichi University of Education, Kariya 448-8542, Japan}

\author[0000-0003-0304-9283]{Tomoya Takiwaki}
\affiliation{Division of Science, National Astronomical Observatory of Japan, Tokyo 181-8588, Japan}

\author[0000-0003-2456-6183]{Kei Kotake}
\affiliation{Faculty of Science, Department of Applied Physics \& Research
  Institute of  Stellar Explosive Phenomena, Fukuoka University, Fukuoka 814-0180, Japan}
\begin{abstract}
  To study properties of magneto-hydrodynamic (MHD) convection and resultant dynamo activities in proto-neutron stars (PNSs),
  we construct a ``PNS in a box'' simulation model with solving compressible MHD equation coupled with a nuclear equation of state (EOS)
  and a simplified leptonic transport. As a demonstration, we apply it to two types of PNS models with different internal structures:
  fully-convective model and spherical-shell convection model. By varying the spin rate of models, the rotational dependence of convection
  and dynamo that operate inside the PNS is investigated. We find that, as a consequence of turbulent transport by rotating stratified convection,
  large-scale structures of flow and thermodynamic fields are developed in all models. Depending on the spin rate and the convection zone depth,
  various profiles of the large-scale structures are obtained, which can be physically understood as steady-state solutions to the ``mean-field''
  equation of motion. Additionally to those hydrodynamic structures, the large-scale magnetic component with $\mathcal{O}(10^{15})$ G is also
  spontaneously organized in disordered tangled magnetic fields in all models. The higher the spin rate, the stronger the large-scale magnetic
  component is built up. Intriguingly, as an overall trend, the fully-convective models have a stronger large-scale magnetic component than that
  in the spherical-shell convection models. The deeper the convection zone extends, the larger the size of the convection eddies becomes. As a result,
  the rotationally-constrained convection seems to be more easily achieved in the fully-convective model, resulting in the higher efficiency of the
  large-scale dynamo there. To gain a better understanding of the origin of the diversity of NS's magnetic field, we need to study the PNS dynamo in
  a wider parameter range.
\end{abstract}
\section{Introduction}
\begin{table*}[ht!]
  \caption{Summary of the simulation runs. The models with prefixes mf \& ms correspond to the models with the
    full-sphere CZ and spherical-shell CZ. The mean quantities $v_{\rm mean}$, $Ro_{\rm mean}$, $\langle \epsilon_{\rm M} \rangle$, $\langle \epsilon_{\rm K} \rangle$
    are evaluated at the saturated state, where $v_{\rm mean} \equiv \langle \langle (v_r - \langle\langle v_r \rangle_\phi\rangle)^2 \rangle_{\rm V}^{1/2}\rangle$
    and $Ro_{\rm mean} \equiv v_{\rm mean}/(2\Omega_0 w_{\rm cz})$ with the width of the CZ, $w_{\rm cz}$, for each model, {\it model}~mf ($w_{\rm cz} = 15$km) or {\it model}~mf ($w_{\rm cz} = 7$km). }
  \centering
\begin{tabular}{  c  c  c  c  c  c  c  c  c  c } \hline\hline
  & $\Omega_0$ [rad/s] & $R_d$ {\small [km]} & $v_{\rm mean}$ {\small [km/s]} & $Ro_{\rm mean}$ & $\langle \epsilon_{\rm K} \rangle$ {\small [erg/cm$^3$]}& $\langle \epsilon_{\rm M} \rangle$ {\small [erg/cm$^3$]} & $\epsilon_{\rm Mm}^{l=1-3}$ {\small [erg/cm$^3$]}&     \\ \hline\hline
mf12p  &$12\pi$   & $17.5$ & $9.18\times 10^2$ & 0.81 & $3.73\times 10^{30} $ & $9.59\times 10^{29} $ & $2.51\times 10^{25}$   \\ \hline
mf60p  &$60\pi$   & $17.5$ & $7.47\times 10^2$ & 0.13 & $2.99\times 10^{30} $ & $8.90\times 10^{29} $ & $5.83\times 10^{26}$   \\ \hline
mf120p &$120\pi$  & $17.5$ & $6.66\times 10^2$ & 0.06 & $2.99\times 10^{30} $ & $8.14\times 10^{29} $ & $4.14\times 10^{27}$  \\ \hline
ms100p &$100\pi$  & $17.0$ & $2.19\times 10^3$ & 0.50 & $6.75\times 10^{30} $ & $1.46\times 10^{30} $ & $3.75\times 10^{25}$  \\ \hline
ms300p &$300\pi$  & $17.0$ & $2.01\times 10^3$ & 0.15 & $5.75\times 10^{30} $ & $1.30\times 10^{30} $ & $5.54\times 10^{25}$  \\ \hline
ms900p &$900\pi$  & $17.0$ & $1.97\times 10^3$ & 0.05 & $4.73\times 10^{30} $ & $1.20\times 10^{30} $ & $1.06\times 10^{26}$  \\ \hline\hline
\end{tabular}
\label{table1}
\end{table*}
Neutron stars (NSs) have the most extreme magnetic field in the universe, typically
trillion, up to quadrillion times more powerful than Earth's. Although, we know, they
are formed as an aftermath of massive stellar core-collapse, the origin of the magnetic
field is still an outstanding issue in astrophysics. Mainly, two possible origins have
been proposed: fossil field and dynamo field hypotheses \citep[e.g.,][]{spruit08,ferrario+15}.
While the former regards it as an inheritance from NS's main sequence progenitor
\citep[e.g.,][]{ruderman72}, the later presumes that it would be generated by some
dynamo processes in newly-born NSs, also known as proto-neutron stars (PNSs)
\citep[e.g.,][hereafter TD93]{ruderman+73,TD93}. Here the PNS is conventionally defined
as a lepton-rich core inside a pseudo-surface with the density of $\rho=10^{11} $ g/cm$^3$ \citep[e.g.,][]{morozova+18,torres+18}. 

One important physical process, which should be examined further in either scenarios, 
is the role of PNS convection. Even if the strong fossil field exists before the collapse,
it should be subjected to vigorous convective motions after the formation of the PNS
\citep[e.g.,][]{epstein79,burrows+88,keil+96}. It is not fully discussed whether the
structure and coherency of the fossil magnetic field are retained in such a tumultuous
situation. At least, independent from the spin rate of the PNS, the turbulent convection
would strongly disturb, locally amplify, and transport the fossil field \citep{nordlund94},
unless the field strength is stronger than the equipartitioned one. The characteristics
of the fossil field acquired before core-collapse thus seems likely to be lost during
the evolution of the PNS. On the flip side, in the dynamo hypothesis, the convective
motion would play a vital role in generating the large-scale magnetic field in the PNS.
Its importance is indisputable. 

The convective dynamo in the PNS is discussed in TD93 theoretically under the modern
scenario of the core-collapse supernova. In the context of the $\alpha$--$\Omega$ dynamo
\citep[e.g.,][]{parker55}, they argue that a large-scale magnetic field with
$\mathcal{O}(10^{15})$ G, is generated when the spin period of the PNS ($\equiv P_{\rm rot}$)
is shorter than $\mathcal{O}(10)$ ms. This constraint comes from the prerequisite for
operating the large-scale convective dynamo \citep[e.g.,][]{moffatt78,krause+80} :
{\it the Rossby number should be smaller than unity}, that is
$Ro \equiv v/(2\Omega l) \lesssim 1$, where the spin rate $\Omega$, the typical velocity
and length-scale $v$ and $l$ with their chosen values $v \simeq \mathcal{O}(10^3)\ {\rm km/s}$
and $l \simeq \mathcal{O}(0.1)\ {\rm km}$, corresponding to the convection velocity and
scale-height expected in the outer region of the PNS. In the ordinary PNS with
$P_{\rm rot} \gtrsim \mathcal{O}(10)$ ms \citep[e.g.,][]{ott2006}, the large-scale magnetic
field does not grow and the small-scale ``patchy'' magnetic structures would prevail.
This is the standard, but still rough, framework they constructed for the PNS dynamo.

The numerical modeling is a powerful tool to refine the PNS dynamo theory. \citet{bonanno+03}
is the first to study the dynamo process in the PNS in the framework of the mean-field
approximation. They have solved a mean-field induction equation with given profiles of
turbulent electro-motive force and differential rotation without solving the development
of the flow field. Then, they showed that the mean-field dynamo process would be effective
for the most of the PNS \citep[see also,][]{bonanno+06}. \citet{rheihardt+05} studies the
ability of the PNS convection to excite the dynamo with considering actual convection 
profiles taken from hydrodynamic simulations of rotating PNSs, and then shows that the
geometrical structures of the velocity fields they employed are well suited to amplify a
seed magnetic field. More recenently, a groundbreaking study for the PNS dynamo was conducted
by \citet{raynaud+20}. They have performed anelastic MHD simulations of the PNS dynamo under
a realistic setup and then confirmed, for the fist time, that the large-scale convective
dynamo successfully occurs in the rapidly-rotating PNS, as predicted by TD93.

Although significant progress has been made in the study of the PNS dynamo with the development
of the numerical method and the improved computing performance, the origin of the diversity of
NS's magnetic fields remains to be solved. ``{\it What physics is responsible for the diversity of NS's magnetic fields ?''.} 
To answer this question, we join the effort of the numerical modeling of the PNS dynamo. The aim
of our study is to give shape to the theory of the PNS convection and dynamo in a more quantitative,
self-consistent manner with the aid of numerical simulation. As a first step towards it, we construct
a ``PNS in a box'' simulation model with solving the compressible magneto-hydrodynamics (MHD) with
a nuclear EOS and a simplified leptonic transport process. Our simulation model has an ability to
study various types of the PNS structure from spherical-shell convection states to full-sphere
convection state. As a demonstration of our newly-developed model, here we study convection and
dynamo processes in two-types of PNSs with different internal structure : one is in a fully-convective
state and the other has a spherical shell convection state. The internal structures of the PNS models
are taken from core-collapse simulations, thus are more or less realistic. The dependence on the spin
rate of the properties of the PNS dynamo is also studied by varying it systematically. 

The outline of this paper is as follows. In \S~2 we present the numerical model we constructed 
and simulation setups. The results obtained by the simulation run for models with different depths
of the convection zone (CZ) and various spin rates are presented, with special focuses on the profiles
of large-scale fields developed in models, in \S~3. The spin-rate dependence of the dynamo activity and its
relationship with the turbulent electro-motive force are discussed in \S~4. We summarize
and discuss the implications of our results in \S~5.
\section{Simulation Model}
\begin{figure*}[ht!]
  \epsscale{1.1}
  \plotone{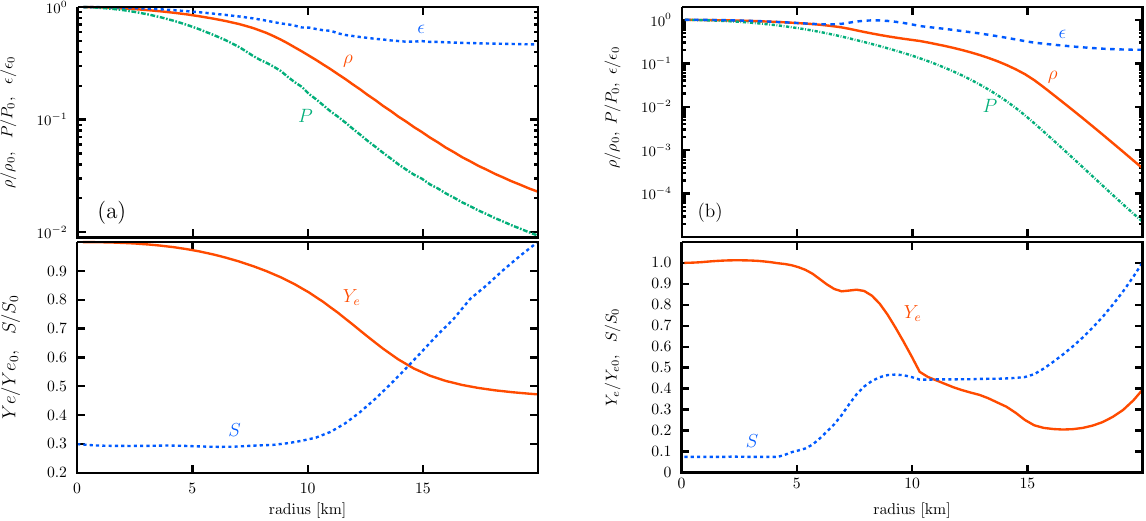}
  \caption{Radial distributions of $\rho$, $P$, $\epsilon$ (top), and $Y_e$, $S$ (bottom)
    for (a) full-sphere convection model ({\it model}~mf) and (b) spherical-shell convection
    model ({\it model}~ms). Normalizations are $\rho_0 = 2.5\times 10^{14}\ {\rm g/cm^3}$,
    $P_0 = 5.86\times 10^{33}\ {\rm dyn/cm^2}$, $\epsilon_0 = 5.85\times 10^{19}\ {\rm erg/g}$,
    $Y_{e0} = 0.35$, and $S_0 = 3.96k_B$ for {\it model}~mf, $\rho_0 = 6.23\times 10^{14}\ {\rm g/cm^3}$,
    $P_0 = 5.54\times 10^{34}\ {\rm dyn/cm^2}$, $\epsilon_0 = 8.15\times 10^{20}\ {\rm erg/g}$,
    $Y_{e0} = 0.27$, and $S_0 = 6.51k_B$ for {\it model}~ms.}
\end{figure*}
It is well known that the width of the CZ in the PNS changes depending not only on the physical properties of the progenitor star
\citep[e.g.,][]{nagakura+19,torres+19} but also on the evolution phase of the PNS \citep[e.g.,][]{keil+96,pons+99,roberts+12,camelio+19}.
Since the CZs with various depths are expected to develop in the interior of the PNS, the numerical model should have an ability
to examine them comprehensively. To understand the origin of the diversity of the NS's magnetic fields, we construct a kind of
``star in a box'' simulation model \citep{freytag+02,dobler+06,kapyla21}.

To cover the whole sphere from the center to pseudo-surface, the PNS is described as a spherical sub-region of radius $R_{\rm PNS}$ of a cubic box
of size $L_{\rm box}^3$ and is solved in the Cartesian grids $(x,y,z)$. The spherical coordinates $(r,\theta,\phi)$ are used for analysis. The baryonic
matter in the box is governed by fully-compressible non-relativistic MHD equations. The general relativistic effects are not taken into consideration in this work.
The leptonic transport which characterizes the PNS is additionally solved under the diffusion approximation. Basic equations are written, in a rotating
reference frame with an angular velocity $\Omega_0$, as
\begin{eqnarray}
   &&\frac{D \rho }{D t} + \rho \nabla\cdot \mbox{\boldmath $v$} = 0 \;, \\
   &&\frac{D \mbox{\boldmath $v$}}{D t} = - 2\Omega_0\mbox{\boldmath $e$}_z\times\mbox{\boldmath $v$}
   - \frac{1}{\rho}\nabla P + \frac{1}{4\pi\rho}\left( \nabla \times \mbox{\boldmath $B$} \right)\times \mbox{\boldmath $B$} \nonumber \\
  &&\ \ \ \ \ \ \ \ \ \ \ \ \ \ \ \ \ \  + \frac{2}{\rho}\nabla\cdot\left(\rho\nu \mbox{\boldmath $S$} \right) + \mbox{\boldmath $g$}
  + \mbox{\boldmath $f$}_{\rm damp} \;, \\
  && \frac{D \epsilon}{Dt} = - \frac{P\nabla\cdot{\mbox{\boldmath $v$}}}{\rho} + 2\nu\mbox{\boldmath $S$}^2 \nonumber \\
  &&\ \ \ \ \ \ \ \ \ \ \ \ \ \ \ \ \ \ + \frac{\eta(\nabla \times \mbox{\boldmath $B$})^2}{4\pi\rho}
  + \frac{\gamma\nabla\cdot\left(\kappa\nabla\epsilon\right)}{\rho} + \dot{\epsilon}_{\rm damp} \;, \\
  &&\frac{\partial \mbox{\boldmath $B$}}{\partial t} = \nabla \times \left( \mbox{\boldmath $v$} \times \mbox{\boldmath $B$}
  - \eta \nabla \times \mbox{\boldmath $B$} \right) \;, \\
  &&\frac{DY_e}{Dt} = \nabla\cdot\left(\xi\nabla Y_e \right) + \dot{Y}_{e,{\rm source}}\;,
\end{eqnarray}
with the strain rate tensor
\begin{equation}
  S_{ij} \equiv (\partial_j v_i + \partial_i v_j -2\delta_{ij}\partial_i v_i/3)/2 \;,
\end{equation}  
where $\epsilon$ is the specific internal energy, $Y_e$ is the lepton fraction, $\gamma$ is the adiabatic index, and the other
symbols have their usual meanings. The viscous, magnetic, heat, and lepton diffusivities are given by $\nu$, $\eta$, $\kappa$ and $\xi$,
respectively. To avoid boundary artifacts, the damping terms $\mbox{\boldmath $f$}_{\rm damp}$ and $\dot{\epsilon}_{\rm damp}$
are added to eqs.~(2) and~(3), which keep $\mbox{\boldmath $v$}$ and $\epsilon$ outside the PNS close to the initial profiles, 
and are given by 
\begin{equation}
  \mbox{\boldmath $f$}_{\rm damp} = - \frac{\mbox{\boldmath $v$}}{\tau_d}f_{\rm ext} \;, \ \ \
  \dot{\epsilon}_{\rm damp} = \frac{\epsilon_{\rm ini} - \epsilon}{\tau_d} f_{\rm ext} \;,
\end{equation}
with
\begin{equation}
  f_{\rm ext} = \frac{1}{2}\left[1 + \tanh \left(\frac{r - R_{\rm d}}{w_t}\right) \right] \;,
\end{equation}
where $\epsilon_{\rm ini}$ is the initial profile of $\epsilon$, $R_{\rm d}$ is the damping radius, $\tau_d$ is the damping time
and $w_t$ is the width of the transition layer between the PNS and the ``buffer'' damping region. For maintaining
the lepton gradient, the source term is added to the lepton transport equation 
\begin{equation}
  \dot{Y}_{e,{\rm source}} = \frac{Y_{e,{\rm ini}} - Y_e}{\tau_{s}} f_{\rm int}
\end{equation}
with
\begin{equation}
  f_{\rm int} = \frac{1}{2}\left[ 1 - \tanh \left(\frac{r - R_{\rm d}}{w_t}\right) \right] \;,
\end{equation}
where $Y_{e,{\rm ini}}$ is the initial profile of $Y_e$, and $\tau_s$ is the forcing time. This is a simple model
for the replenishment of the lepton via the energy conversion from the gravity to the neutrino radiation inside
the PNS during its cooling time\footnote{Our PNS dynamo simulations focus on the timescales of less than ``300 ms'', which is shorter than typical
thermodynamic evolution time of the PNS (e.g., $\sim 1000$ ms in \citet{scheck+06} for the standard case). On such timescale, the behavior
of neutrino transport is expected to be similar to (not so different from) that in the condition adopted as our initial setup. Hence,
we models the lepton-driven convection by adding the forcing term, that can maintain the initial profile of the lepton fraction moderately,
to the lepton transport equation.}. To close the system, we employ the EOS by \citet{LS91} with a compressibility
modulus of $K = 220$ MeV.

We set up two-types of initial equilibrium models as typical examples for our newly-developed code : {\it model}~mf is based
on a post-bounce core (about $100$ ms after the core bounce) from a hydrodynamic simulation of ``rotating'' core-collapse of
$15M_\odot$ progenitor and 
{\it model}~ms is based on a post-bounce core (about $600$ ms after the bounce) of ``non-rotating'' core-collapse of $11.2M_\odot$
progenitor. In both models, the shock wave has reached $\sim 200$ km, and the PNSs are settled into a quasi-hydrostatic state.
The hydrodynamic variables and gravitational potential within $0\le r \le 20$ km ($\equiv  L_{\rm box} /2$) are extracted,
and then the PNS is reconstructed with a 2nd-order interpolation method in the calculation domain ranging
$-L_{\rm box}/2 \le x, y, z \le L_{\rm box}/2$. 
\begin{figure}[ht!]
  \epsscale{1.1}
  \plotone{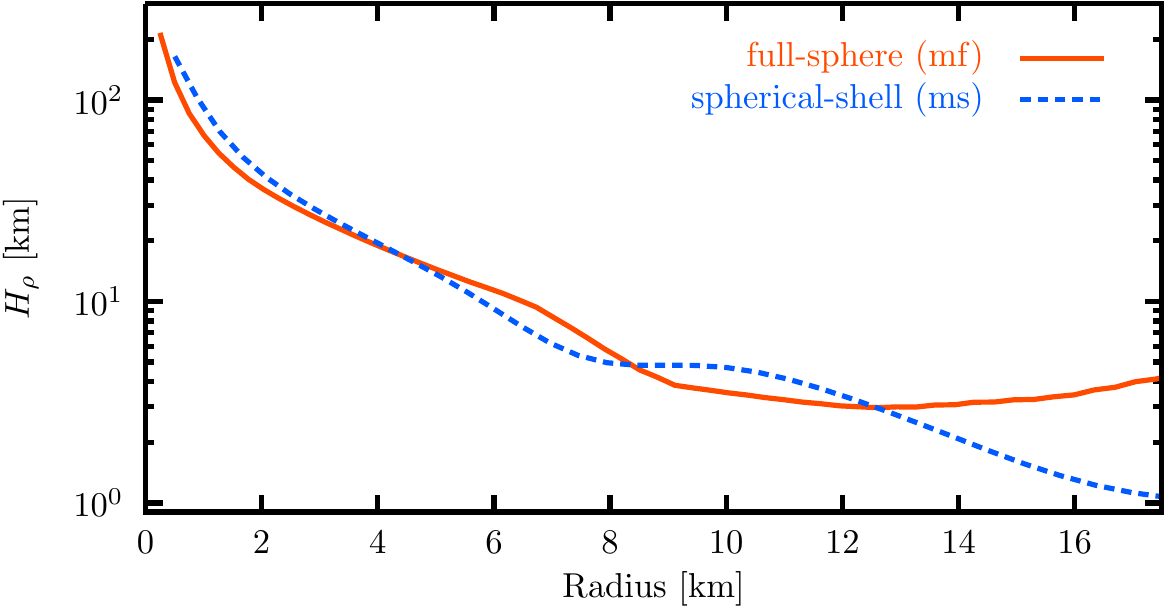}
  \caption{Radial distribution of the density scale-height ($H_\rho \equiv -dr/d\ln\rho$) for {\it model}~mf (red-solid)
    and {\it model}~ms (blue-dashed).}
\end{figure}

Shown in Figure 1 is the initial profile of the hydrodynamic variables for (a) {\it model}~mf and (b) {\it model}~ms. While the profiles of
$\rho$, $P$, and $\epsilon$ are shown in top panels, $Y_e$ and $S$ (entropy) are in bottom panel in each figure. The radial profile of the
density scale-height $H_\rho$ for each model is also shown in Figure~2. There is no significant difference in the profile of $H_\rho$ between
two models. However, there are remarkable differences in the profiles of $Y_e$ and $S$ between models. For {\it model}~mf, while the negative
lepton gradient which lies in the inner part of the PNS powers the lepton-driven convection \citep[e.g.,][]{epstein79,keil+96}, the outer
region where $r \gtrsim 15$ km is stable to the convective instability based on the Ledoux criterion \citep[e.g.,][]{ledoux47,KW90}. In contrast,
for {\it model}~ms, the possible site for the lepton-driven convection is confined in the layer in the range $7.5$km $\lesssim r \lesssim 15$ km,
while the inner core and outer envelope are convectively stable due to the positive entropy gradient. In this paper, to explore the response of
the PNS convection and dynamo to the spin rate, we vary the magnitude of $\Omega_0$ in two models while keeping the background hydrostatic state
unchanged. We investigate the cases with $\Omega_0 = 12\pi, 60\pi$ and $120\pi$ rad/s for {\it model}~mf and the cases with
$\Omega_0 = 100\pi, 300\pi$ and $900\pi$ rad/s for {\it model}~ms, respectively. See the summary of the simulation run in Table~1. 

Since the outer convectively stable layer has less impact on the hydrodynamics of the PNS during the evolution time of interest
($\mathcal{O}(100)$ ms), the damping radius is chosen as $R_{\rm d} = 17.5$ km so that the pseudo-surface of the PNS would be $R_{\rm PNS}\simeq 15$ km.
We connect the PNS to the outer buffer region through the transition layer with $w_t = 0.05R_{\rm PNS}$. The forcing time for the lepton is assumed
to be constant inside the PNS and an order of magnitude shorter than the typical convective turn-over time, that is $\tau_s = 0.1\tau_{\rm cv}$,
where $\tau_{\rm cv} \equiv l_{\rm sh}/v_{\rm cv}$ with the typical convection velocity $v_{\rm cv} = 10^8\ {\rm cm/s}$ and the typical scale-height
$l_{\rm sh}=10^5\ {\rm cm}$. With this value, we can keep the profile of $Y_e$ close to, but slightly deviate from, the initial state. To reduce
the boundary artifacts as much as possible, we adopt a short damping time of $\tau_d = 0.1\tau_s$. We choose, as a first step, uniform
diffusivities of $\nu = \eta = \kappa = \xi = 10^{11} {\rm cm^2/s}$ inside the PNS. While $\nu$, $\kappa$ and $\xi$ assumed here are within the
expectations in the PNS\footnote{In the PNS below the neutrinosphere, the neutrino diffusion process controls the magnitude of the diffusion
coefficients, except the magnetic diffusivity (Thompson \& Duncan 1993). }, $\eta$ is far from realistic value
($\eta_{\rm PNS} \sim 10^{-5}\ {\rm cm^2/s}$) \citep[e.g.,][]{rheihardt+05,masada+07} 
because of a numerical limitation as is so often the case with the planetary and stellar dynamo simulations \citep[e.g.,][]{jones11,brun+17}.  

The governing equations are solved by the second-order Godunov-type finite-difference scheme which employs an approximate MHD Riemann solver
\citep[see][for details]{sano+99,masada+12,masada+15}. The calculation domain is resolved by $N^3 = 256^3$ grid points. 
After a random small ``seed'' magnetic field with the amplitude $|\delta B| < 10^9$ G is introduced into the CZ of the PNS, the
calculation is started by adding a small perturbation to the initial pressure distribution for both models. Note that, in this paper, the
simulation is terminated when the calculation time exceeds roughly $250$ ms, with taking into account that the structure of the PNS can
change in about a few hundred milli-seconds as the neutrino radiation proceeds \citep[e.g.,][]{keil+96,scheck+06, nagakura+19}.

In the following, to examine the convective and magnetic structures in detail, we define the following three averages of an arbitrary function
$h(r, \theta,\phi)$ \footnote{The function $h(r,\theta,\phi)$ is an arbitrary function used only to explain three different averaging methods (volume, longitudinal and spherical averages) adopted in our analysis and their notations in this paper. In the main body, $h$ is replaced by the velocity $v_i$ or the magnetic field $B_i$ ($i=x,y,z$).}.\\
The volume average over the entire CZ:
\begin{equation}
\langle h \rangle_{\rm V} \equiv \int h(r,\theta,\phi) {\rm d}V_{\rm cz} /\int {\rm d}V_{\rm cz} \;,
\end{equation}
where $V_{\rm cz}$ is the volume of the CZ. \\
The longitudinal average:
\begin{equation}
\langle h \rangle_\phi = \frac{1}{2\pi} \int_{-\pi}^{\pi} h(r,\theta,\phi) {\rm d}\phi \;.
\end{equation}\\
The spherical average:
\begin{equation}
\langle h \rangle_s = \frac{1}{4\pi} \int_{-1}^{1} \int_{-\pi}^{\pi} h(r,\theta,\phi){\rm d}\cos\theta{\rm d}\phi\;.
\end{equation}
The time-average of each spatial mean is denoted by additional angular brackets, such as $\langle\langle h \rangle_\theta\rangle$. 

\section{Results}
\subsection{Temporal Evolution and Convection Profiles}
\begin{figure}[ht!]
  \epsscale{1.1}
  \plotone{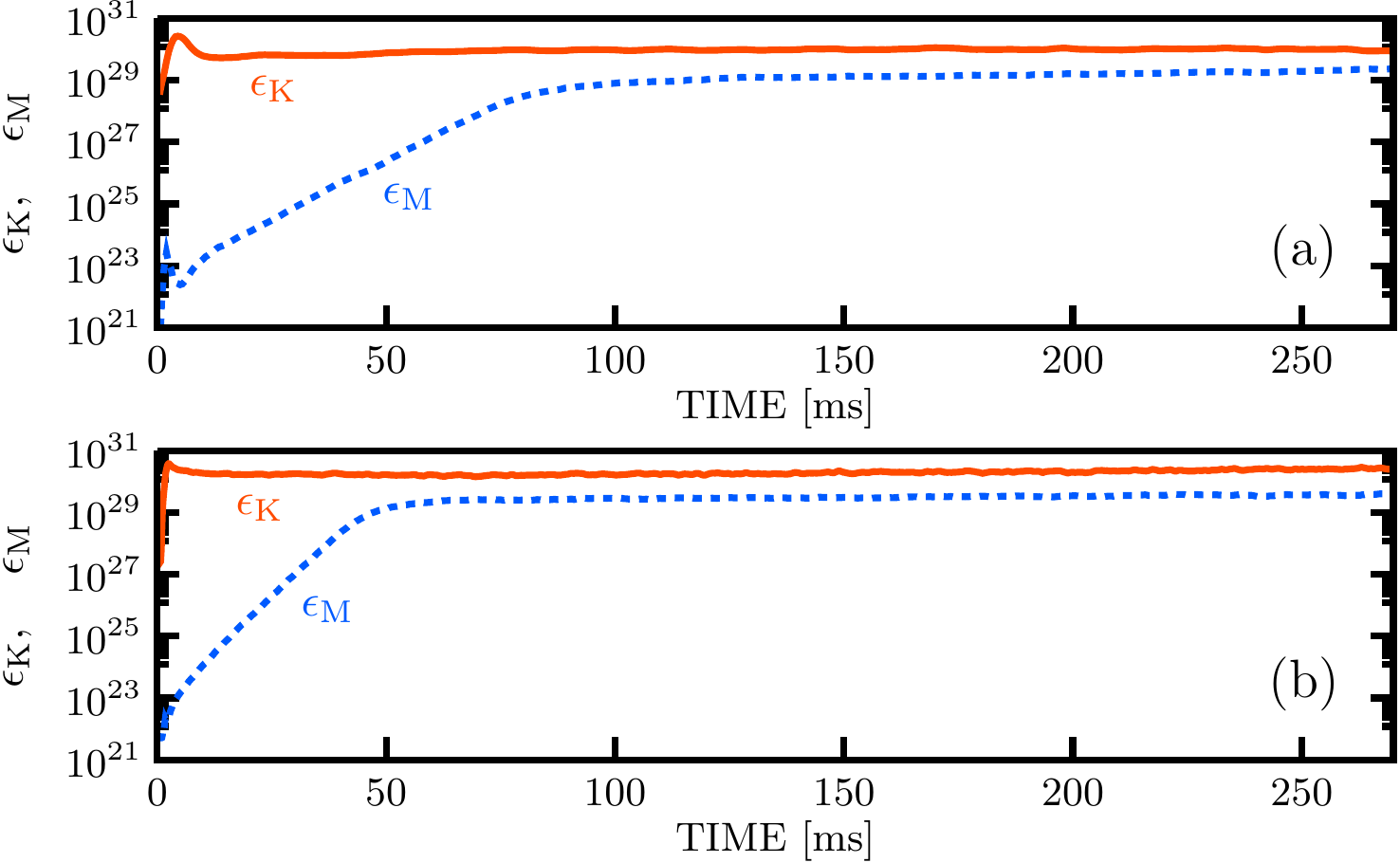}
  \caption{Temporal evolution of $\epsilon_{\rm K}$ (red-solid) and $\epsilon_{\rm M}$ (blue-dashed) for (a) {\it model}~mf and (b) {\it model}~ms.}
\end{figure}
\begin{figure}[ht!]
  \epsscale{1.0}
  \plotone{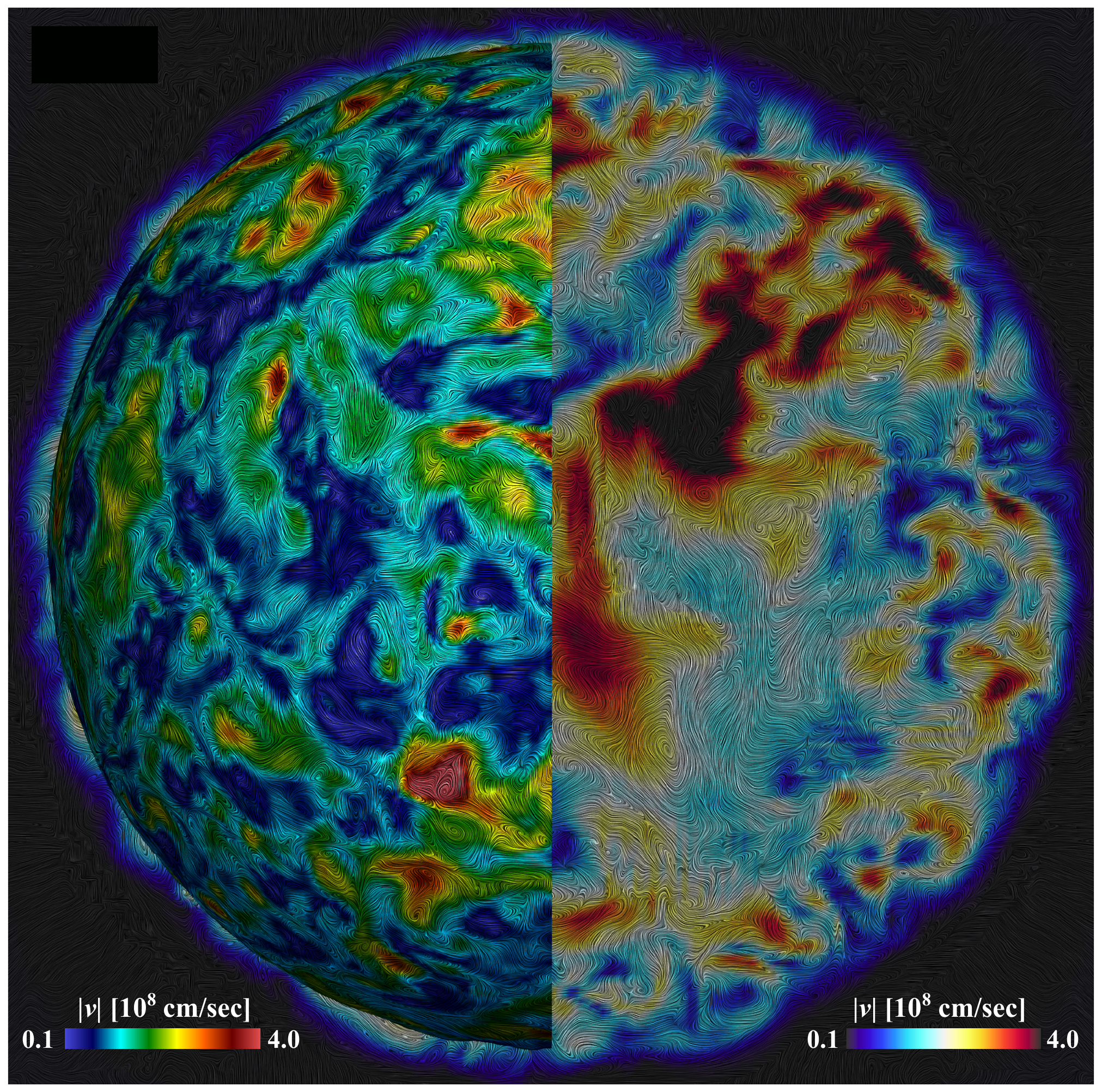}
  \caption{LIC visualizations of $|\mbox{\boldmath $v$}|$ at $r = 15$km (left hemisphere) and meridional cutting plane (right hemisphere) when
    $t = 230$ ms for {\it model}~mf12p as a demonstration. Normalization unit is $10^8\ {\rm cm/s}$. Red (blue) tone denotes higher (lower) convection
    velocity.}
\end{figure}
The rough sketch of the simulation results is summarized. The typical temporal evolution of the volume-averaged kinetic and magnetic energies
($\epsilon_{\rm K}$ and $\epsilon_{\rm M}$) for (a) {\it model}~mf (mf12p) and (b) {\it model}~ms (ms100p) is shown in figure~3, where
$\epsilon_{\rm K} \equiv \langle \rho {\bm v}^2/2 \rangle_{\rm V}$ and $\epsilon_{\rm M} \equiv \langle {\bm B}^2/(8\pi) \rangle_{\rm V}$. The volume-average is
taken over the entire CZ. When the simulation proceeds, the lepton-driven convection begins to grow and then both kinetic and magnetic
energies are amplified in the PNS. The kinetic energy saturates within $\sim 50$ ms. At the saturated stage, the turbulent convection motion dominates
the PNS as shown in figure~4 where the structure of the convection on the PNS surface (left hemisphere) and the meridional cutting plane (right
hemisphere) for {\it model}~mf (mf12p) is visualized as a demonstration. In contrast, the magnetic energy gradually increases and reaches a quasi-steady
state after $t = 150\sim 200$ ms in both models. Since the convective turn-over time is longer in {\it models}~mf than in {\it models}~ms due to the
larger scale-height in the deeper CZ, it takes longer time for the magnetic energy to be amplified. The saturation level of the magnetic
energy is below that of the convective kinetic energy, suggesting that the MHD dynamos operated in our models are in a turbulent regime rather
than a magneto-strophic regime \citep[e.g.,][]{brun+17,raynaud+20}. See Table~1 for more details on the temporal means of $\epsilon_{\rm K}$ and
$\epsilon_{\rm M}$ for each model. 

\begin{figure}[ht!]
  \epsscale{1.1}
  \plotone{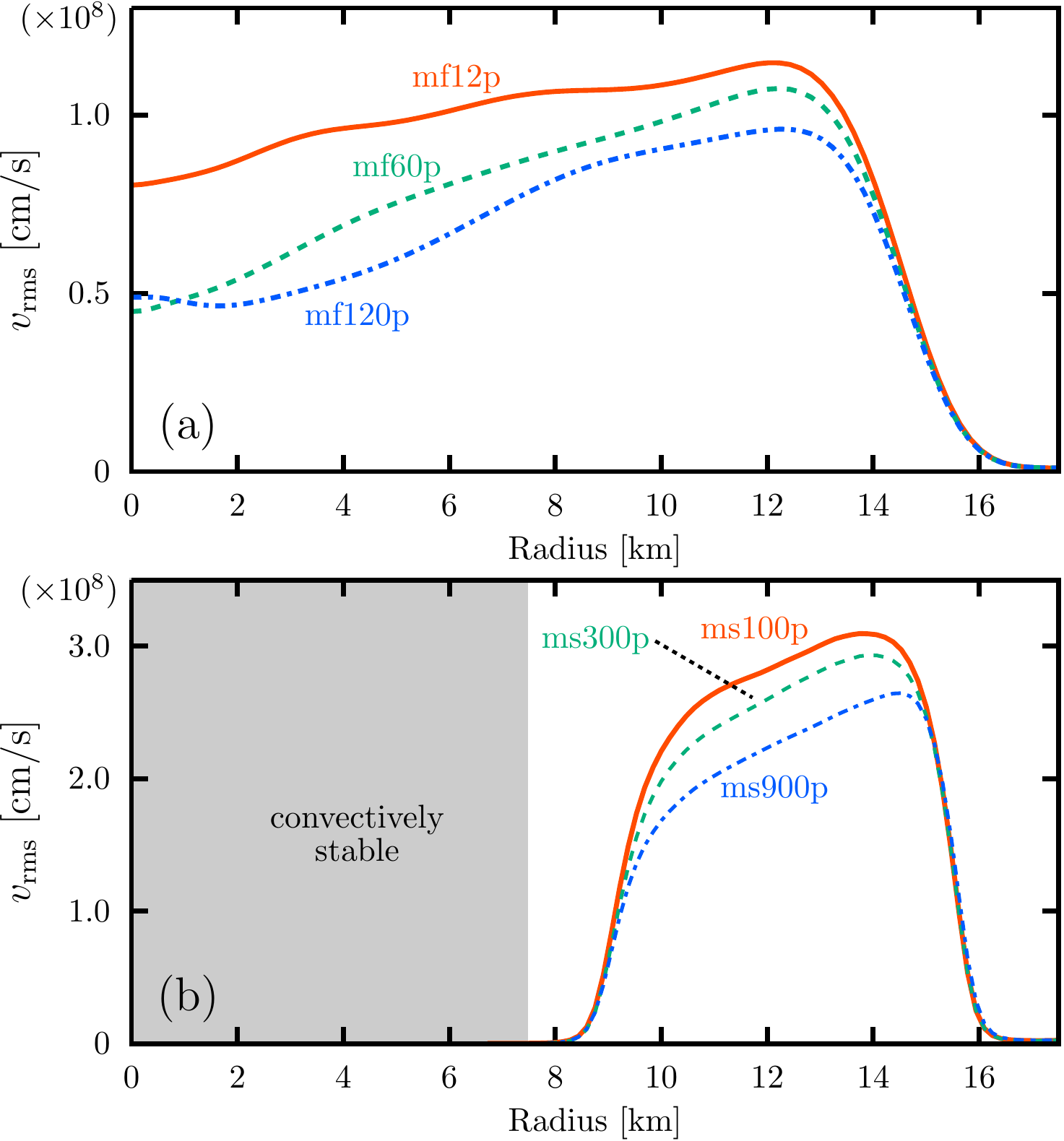}
  \caption{Radial profiles of $v_{\rm rms}$ for (a) {\it models}~mf and (b) {\it models}~ms at the equilibrated stage.
    The different line types correspond to the models with different spin rates in each panel.}
\end{figure}
The radial profiles of $v_{\rm rms}$ at the saturated stage for models (a) mf and (b) ms are shown in figure~5, where 
$v_{\rm rms}$ is the RMS of the perturbed radial velocity defined by 
$v_{\rm rms} \equiv \langle \langle (v_r - \langle\langle v_r \rangle_\phi\rangle)^2 \rangle_{s}^{1/2}\rangle$. The different line types correspond to the models
with different spin rates. The shaded region in panel~(b) denotes the convectively-stable region.

Overall, the convection velocity is an order of $10^{8}\ {\rm {cm/s}}$ although it is a few times higher in {\it models}~ms
than in {\it models}~mf because of the larger thermal energy stored in {\it model}~ms at the initial setup stage (See, fig.~1). 
Since the rotational constraint, caused by the Coriolis force, on the convective motion becomes stronger in the model with
the higher spin rate, the convection velocity decreases as the spin rate increases, which can be seen in each model (See, Table~1). 

As a corollary of the turbulent convective motion regulated by the rotation, large-scale flow, thermodynamic and magnetic structures
are built up in the PNS. In the following, we will discuss in detail the results obtained with each model. 

\subsection{{\it Models}~mf : Full-sphere Convection Models}
\subsubsection{Mean Flow and Thermodynamic Fields}
\begin{figure*}[ht!]
  \epsscale{0.85}
  \plotone{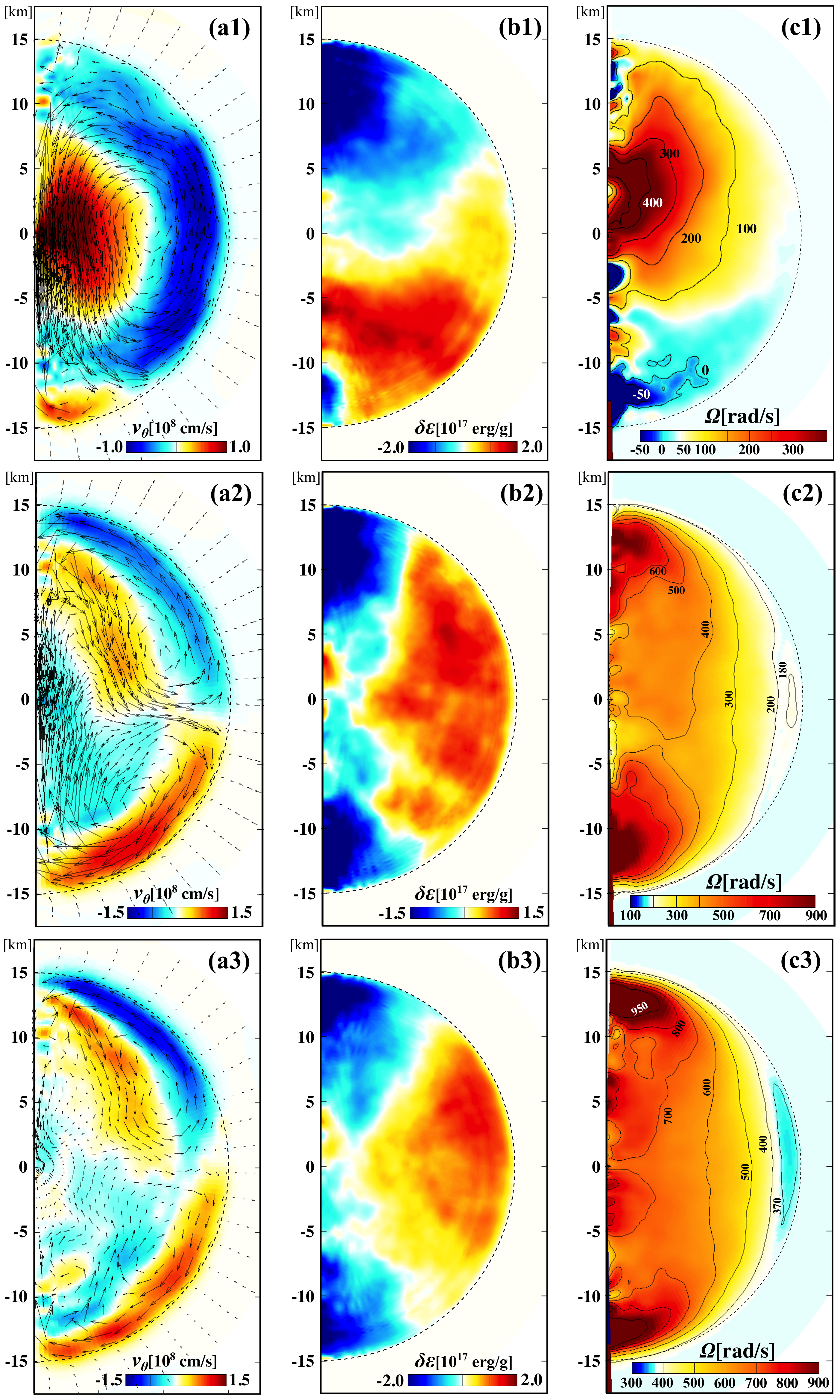}
  \caption{Meridional distributions of (a) $\langle \langle v_\theta \rangle_\phi \rangle$, (b) $\delta \epsilon$ and (c) $\Omega$ for {\it model}~mf.
    The top, middle and bottom panels are corresponding to {\it models}~mf12p, mf60p, and mf120p, respectively. 
    In panel~(a), the streamlines are overplotted with an arrow length proportional to the flow velocity. 
    In panel~(c), the region rotating with the reference frame $\Omega_0$ is shown by white color. The iso-rotation
    contours are also overplotted. }
\end{figure*}
Meridional distributions of (a) $\langle \langle v_\theta \rangle_\phi \rangle$,
(b) $\delta \epsilon \equiv \langle\langle \epsilon - \langle\langle{\epsilon}\rangle_s\rangle\rangle_\phi\rangle$
and (c) $\Omega \equiv \langle\langle v_\phi \rangle_\phi\rangle/r\sin\theta + \Omega_0$ are shown in Figure~6. The overplotted arrows in panel~(a) are meridional
velocity vectors with arbitrary amplitudes. Panels~(a1)--(c1) are for mf12p, (a2)--(c2) are for mf60p, and (a3)--(c3) are for mf120p,
respectively.

From the symmetry point of view, the large-scale flows seen in {\it models}~mf can be divided into two types :
(i) a single-cell meridional circulation with a north-south anti-symmetric differential rotation (mf12p), and (ii) a double-cell meridional
circulation (single-cell per hemisphere) with a cylindrical differential rotation (mf60p and mf120p). While the dipole dominance in
$\delta \epsilon$ is accompanied with the flow pattern (i), it can be seen that the quadrupolar dominance in $\delta \epsilon$ is
developed in conjunction with the pattern (ii). 

The single-cell counter-clockwise profile with circulating between northern and southern hemispheres, seen in fig.6(a1),
should be due to the very slow spin of mf12p. As shown in \citet{chandra61} (see \S~59), the dipole dominance of the convective flow
is a natural topological result of the full-spherical convection domain. See \S~3.3.1 for related discussion. Since the northern and
southern hemispheres are dominated respectively by cool (and fast) down-flow and warm (and slow) up-flow, the large-scale coherent
circulation in-between them is formed, resulting in the anti-symmetric profile of $\delta \epsilon$ with respect to the equator (see fig.6(b1)).
The difference of $\delta \epsilon$ between hemispheres is averagely $4\times 10^{17} \ {\rm erg/g}$ which provides $10^{50}-10^{51}\ {\rm erg}$
when considering the total mass of the PNS. 

The profile of $\Omega$ for mf12p (fig.6(c1)) should be determined to retain a quasi-steady convective state:
the production of vorticity by the baroclinic term ($\propto \partial \overline{\epsilon}/\partial \theta $) 
should be balanced mainly with the production of relative vorticity by the stretching ($\propto \partial \Omega/\partial z$), that is,  
the thermal wind balance 
\begin{equation}
  \frac{\partial \omega}{\partial t} = r\sin\theta \frac{\partial\Omega^2}{\partial z}
  - \frac{g}{\gamma \bar{\overline{\epsilon}}}\frac{\partial \overline{\epsilon}}{\partial \theta}
  \cdot\cdot \cdot = 0\;, 
\end{equation}
should be maintained to retain the steady state, where $\omega$ is the vorticity and $\overline{\epsilon}$ is the time and
azimuthally-averaged specific internal energy. Note that this equation can be obtained as the meridional component of the
mean-field equation of motion \citep[see, e.g.,][]{pedlosky82,masada11}. 

Eq.(14) predicts that the latitudinal variation of $\overline{\epsilon}$ produces a change of $\Omega$ in the $z$-direction.
Since $\partial \overline{\epsilon}/\partial\theta \gtrsim 0$ over the entire CZ in mf12p, the spin rate should increase in the $z$-direction
to satisfy eq.(14). This is qualitatively consistent with what we observe in mf12p, i.e.,  differential rotation with a north-south
anti-symmetry, which progrades in the north and retrogrades in the southern hemisphere of the PNS (see fig.6(c1)). The overall mean properties
of the hydrodynamics seen in mf12p are analogous to those in \citet{brun+09} (see their model labeled RG2) though their simulated object is the
red giant which is different from us. 

The profile of the meridional flow seen in {\it models} mf60p and mf120p consists of a double-cell circulation (single-cell per hemisphere (see fig.6(a2) or (a3)).
While it is poleward at the upper CZ, it reverses to an equatorward direction at deeper depths. In such a case, the polar and equatorial
regions are dominated respectively by cool down-flow and warm up-flow, naturally resulting in the quadrupolar structure of the thermodynamic
field seen in the profile of $\delta \epsilon$ (fig.6(b2) or (b3)). This type of circulation flow is often seen in convection simulations of slowly-rotating
solar-type stars with thin CZs \citep[e.g.,][]{mabuchi+15}. The driving mechanism of such a double-cell circulation (single-cell per hemisphere)
will be discussed in detail in \S~3.3.1 because the similar profile can be observed even in {\it models}~ms. 

A remarkable property of the differential rotation achieved in mf60p and mf120p is that it is almost invariant along the rotation axis,
i.e., $\partial\Omega/\partial z \simeq 0$ (see fig.6(c2) or (c3)). It is well-known that, according to the Taylor-Proudman constraint,
the fluid parcel tends to move and thus its velocity to be uniform along the rotation axis in the system with sufficiently large Coriolis force  
\citep[][]{pedlosky82,kitchatinov+95,brun+02,miesch+06,masada11}. The latitudinal variation of $\Omega$ is the anti-solar type, that is
the spin rate is lower at the equator than in the polar regions. The anti-solar type differential rotation is also a characteristic feature
of the relatively slowly spinning system. It should be noted that the overall mean flow properties in mf60p and mf120p are also analogous to
those in \citet{brun+09} (see their model labeled RG1) where the deep core convection in the red-giant is discussed. 

\subsubsection{Magnetic Field : Dynamo Activities}
\begin{figure}[ht!]
  \epsscale{1.05}
  \plotone{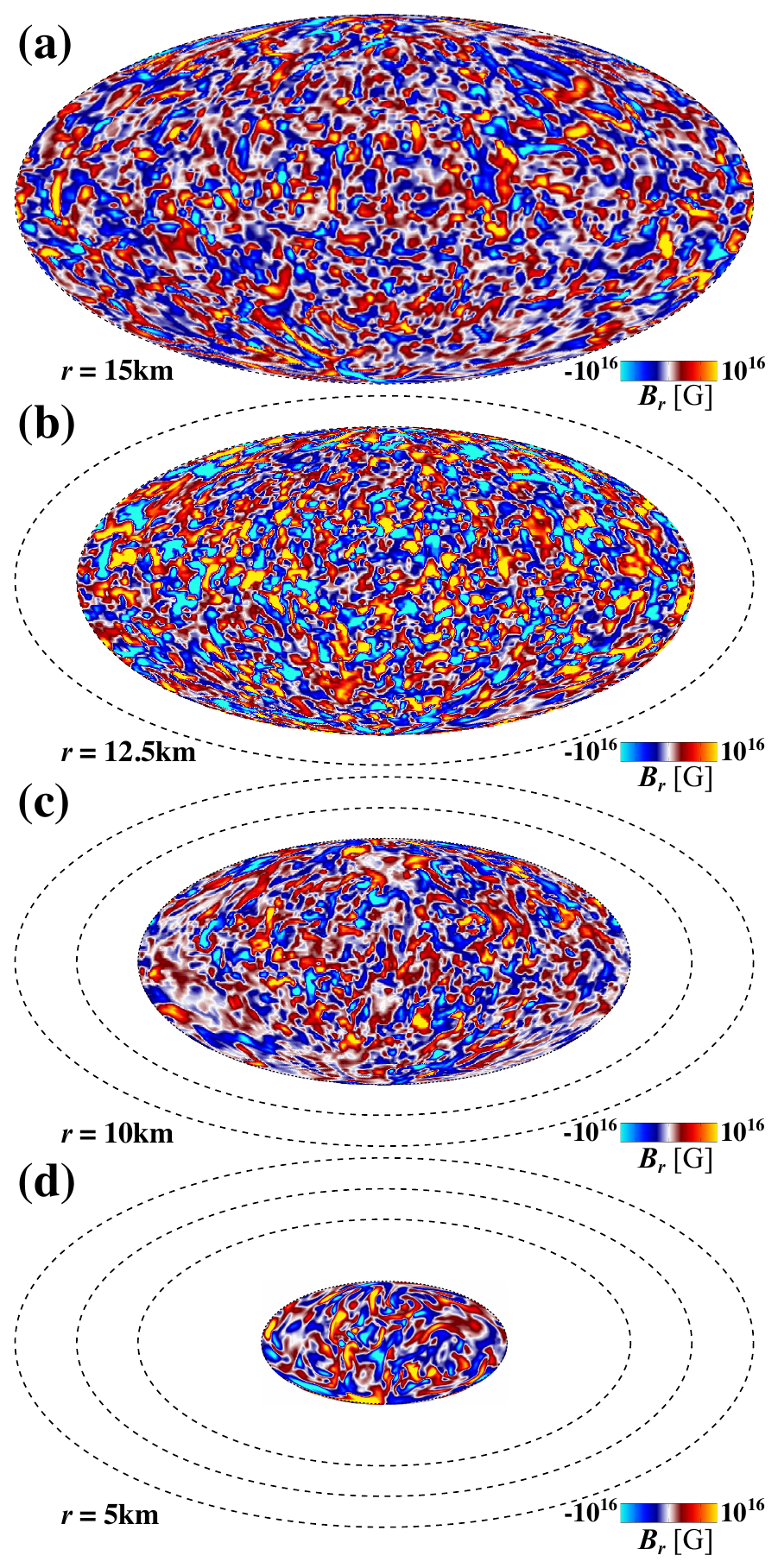}
  \caption{Distributions of the radial component of the magnetic field on spherical
    surfaces at sampled radii $r = 15$km, $12.5$km, $10$km, and $5$km when $t = 230$ ms
    in the Mollweide projection for {\it model}~mf (mf12p) as a reference.}
\end{figure}
\begin{figure*}[ht!]
  \epsscale{0.8}
  \plotone{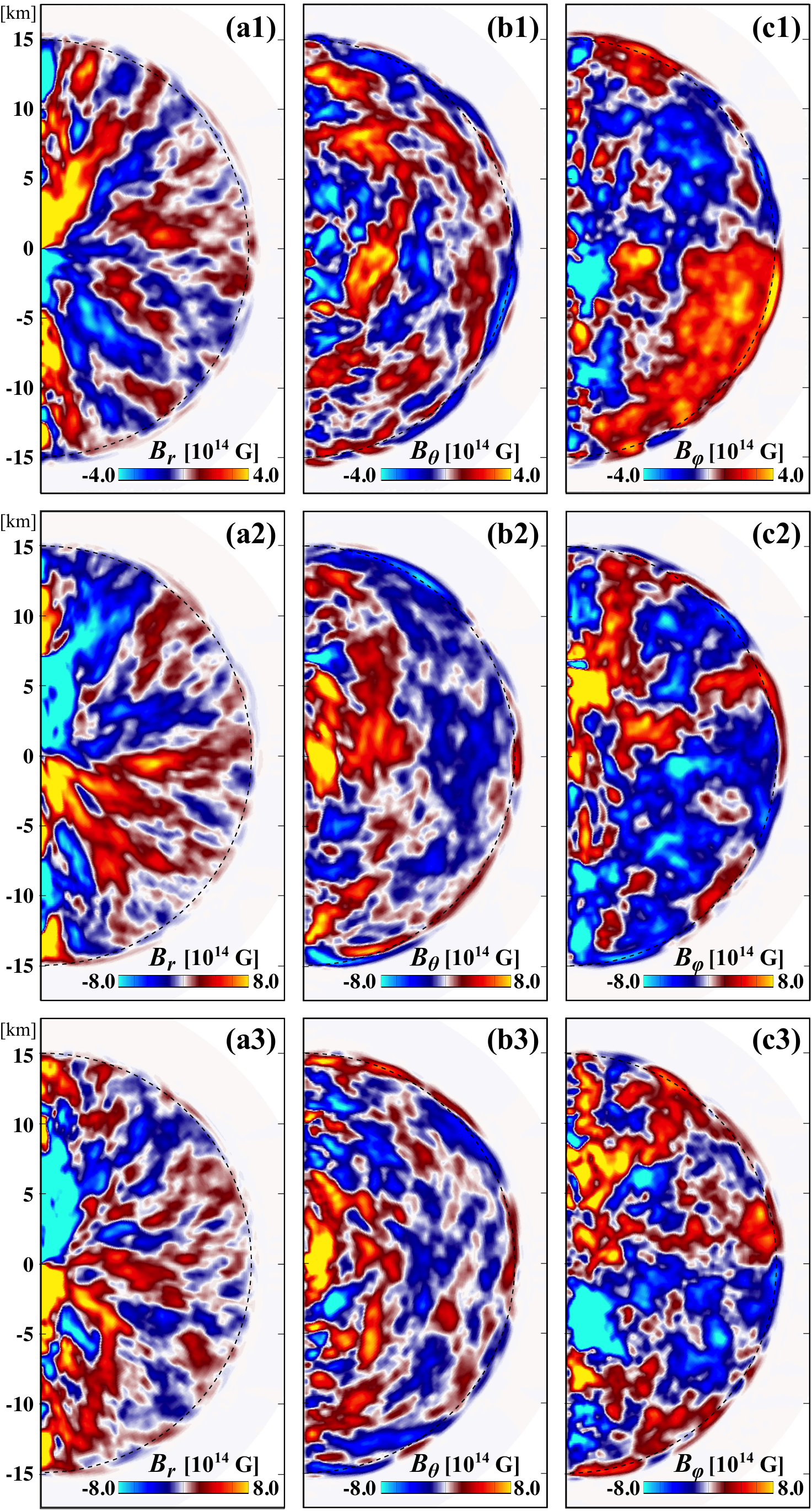}
  \caption{Meridional distributions of large-scale magnetic components, (a) $\langle\langle B_r \rangle_\phi\rangle$,
    (b) $\langle\langle B_\theta \rangle_\phi\rangle$ and (c) $\langle\langle B_\phi \rangle_\phi\rangle$ for {\it model}~mf.
    The top, middle and bottom panels are corresponding to {\it models}~mf12p, mf60p, and mf120p, respectively.
    The time average is taken over the duration $220 \le t \le 240$ ms.}
\end{figure*}
\begin{figure*}[ht!]
  \epsscale{1.15}
  \plotone{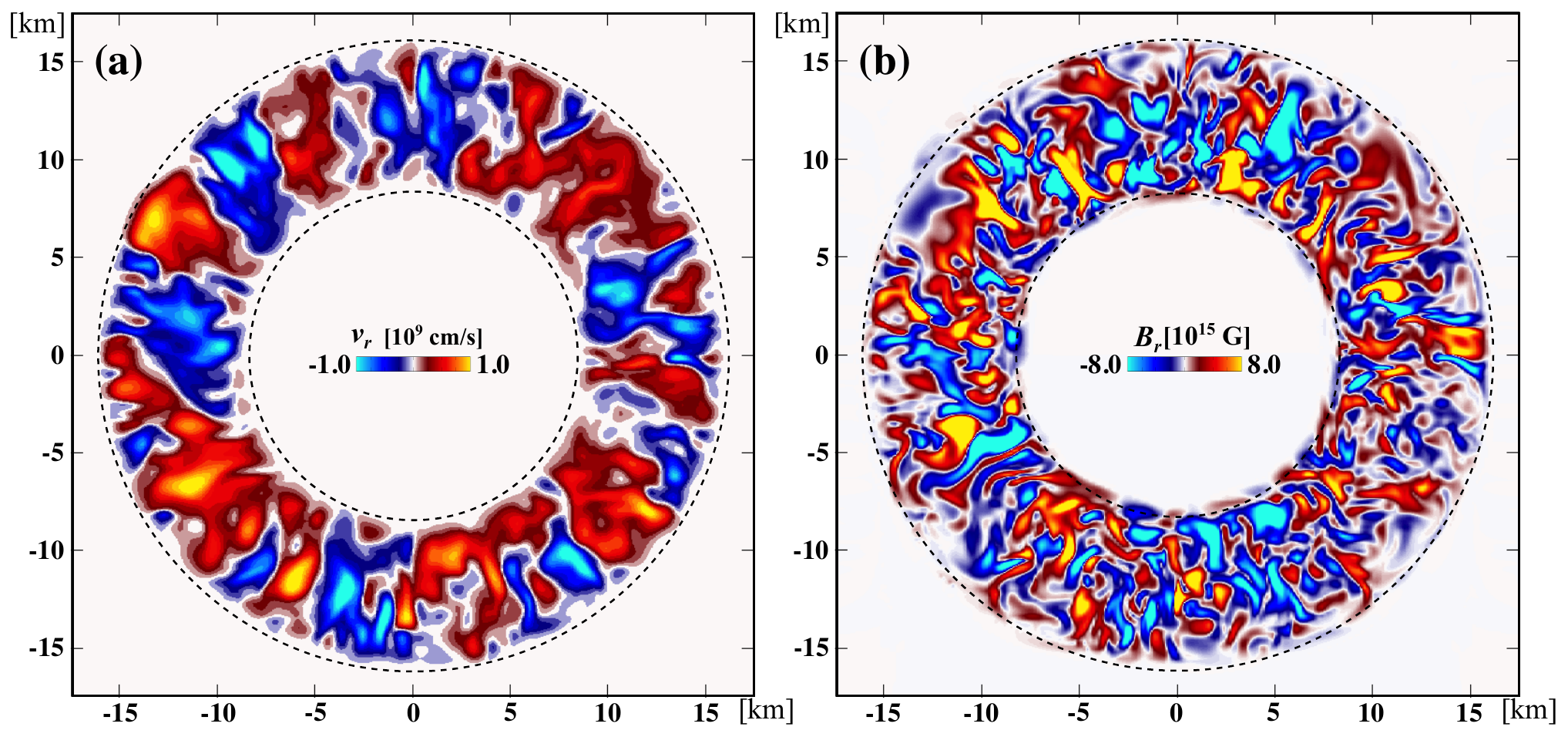}
  \caption{Instantaneous snapshots for the distributions of (a) $v_r$ and $B_r$ when $t = 220$ ms for ~{\it model}~ms (ms100p) at the meridional cutting plane.
  The red (blue) tone denotes the positive (negative) values.}
\end{figure*}
The magnetic field which is amplified by the PNS convection shows a complicated structure mixed with the turbulent and large-scale components.
Primarily, the turbulent convective motion produces small-scale intense magnetic fields. Figure~7 shows, as a demonstration, the distribution of the
radial component of the magnetic field on spherical surfaces at different depths for mf12p in the Mollweide projection. The red (blue) tone denotes
the positive (negative) value of the field strength. It is found that the turbulent component of the magnetic field becomes predominant inside the PNS.
It is amplified via small-scale convective dynamo \citep[e.g.,][]{cattaneo99,schekochihin+04} and finally reaches to the strength of
$\mathcal{O}(10^{16})$ G, which contains the energy of about $30$ \% of the convective kinetic energy at the saturated stage
(see, fig.3 and Table 1). Turbulent magnetic components show similar characteristics in other full-sphere convection models, i.e., mf60p and mf120p. 

In such a haystack of turbulent magnetic components, the large-scale magnetic structure is spontaneously organized in {\it models}~mf. 
Shown in figure~8 is the meridional distributions of large-scale magnetic components, (a) $\langle\langle B_r \rangle_\phi\rangle$,
(b) $\langle\langle B_\theta \rangle_\phi\rangle$ and (c) $\langle\langle B_\phi \rangle_\phi\rangle$ for {\it models}~mf. The top, middle and bottom panels
are corresponding to {\it models}~mf12p, mf60p, and mf120p, respectively. The time average is taken over the duration $220 \le t \le 240$ ms.
The red (blue) tone denotes the positive (negative) magnetic filed strength. 

Commonly, the global structure of the mean magnetic component exhibits a dipole dominance in these models. It can be found that the large-scale poloidal
component, rooted deep in the central part of the PNS, shows a strong dipole symmetry, while it is less coherent in the outer part of the sphere. The
strength of it reaches averagely $\mathcal{O}(10^{14})$ G (locally exceeds $10^{15}$ G), which is compatible with that expected in the strongly-magnetized
NSs, so-called ``magnetars'' \citep[e.g.,][and references therein]{turolla+15,enoto+19}. Additionally to the poloidal component, a large-scale toroidal magnetic
component is also built up mainly in the outer part of the PNS, especially in the relatively slowly-spinning models (mf12p and mf60p). It is roughly
anti-symmetric with respect to the equator and also has an average strength of $\mathcal{O}(10^{14})$ G, which is a bit weaker than the poloidal component. 

The large-scale magnetic component observed in {\it models}~mf would be the self-organized structure as a natural outcome of the symmetry breaking
forced by the NS's spin. Although the successful large-scale dynamo has been observed in a lot of solar, stellar, and planetary MHD convection 
simulations \citep[e.g.,][]{ghizaru+10,kapyla+12,masada+13,fan+14,hotta+16}\citep[See,][for review]{charbonneau20}, our intriguing finding here is
that the large-scale field can be organized even in a relatively slow rotation regime than that predicted by TD93. The dependence of the larges-scale magnetic
field on the spin rate and the mechanism for the large-scale dynamo observed in our simulation models are discussed comprehensively in~\S~4. 

\subsection{{\it Models}~ms : Spherical-shell Convection Models}
\subsubsection{Mean Flow and Thermodynamic Fields}
\begin{figure*}[ht!]
  \epsscale{0.85}
  \plotone{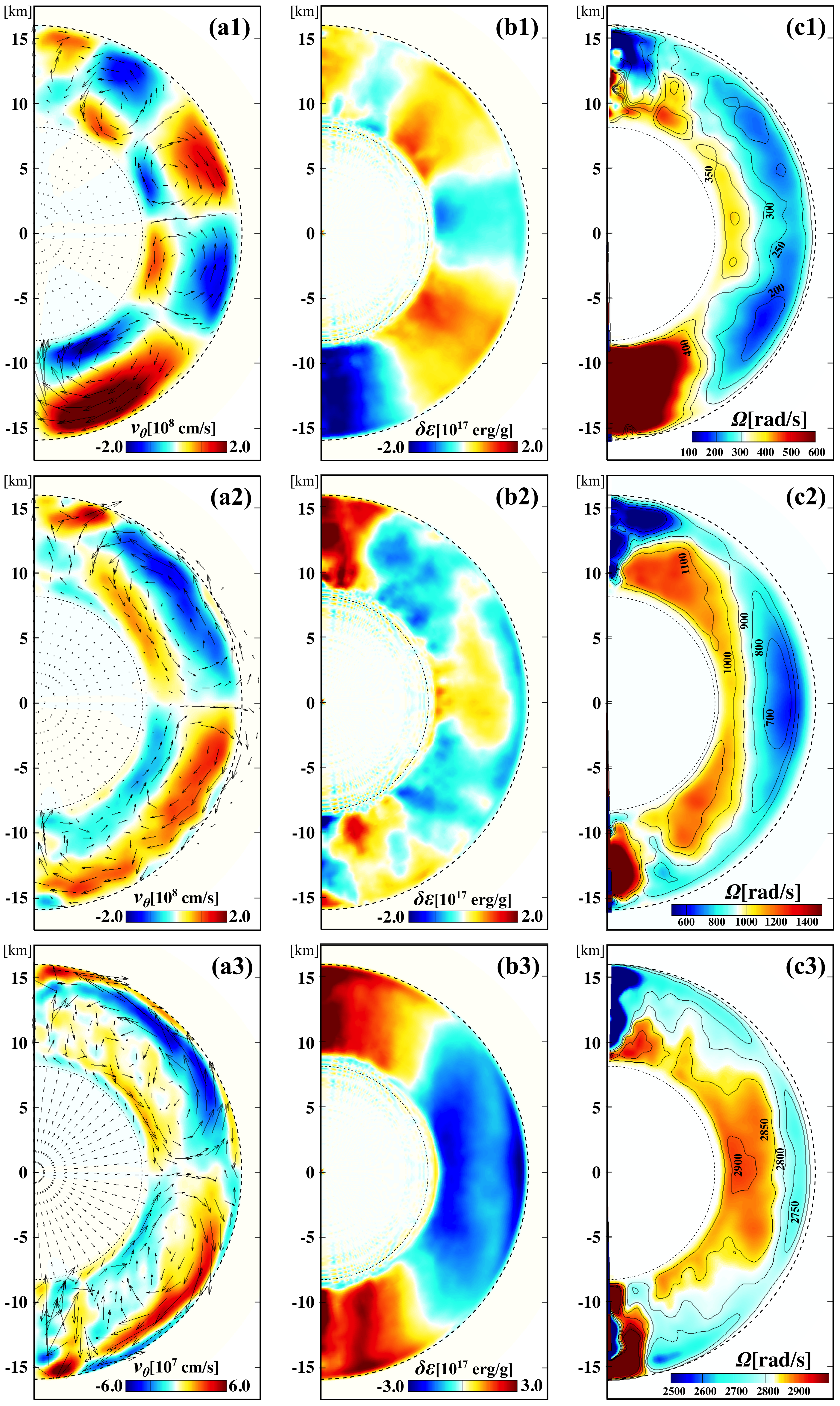}
  \caption{Meridional distributions of (a) $\langle \langle v_\theta \rangle_\phi \rangle$,
    (b) $\delta \epsilon \equiv \langle\langle \epsilon - \langle\langle{\epsilon}\rangle_s\rangle\rangle_\phi\rangle$
    and (c) $\Omega \equiv \langle\langle v_\phi \rangle_\phi\rangle/r\sin\theta + \Omega_0$ for {\it models}~ms.
    The top, middle and bottom panels are corresponding to {\it models}~ms100p, ms300p, and ms900p, respectively. 
    In panel~(a), the streamlines are overplotted with an arrow length proportional to the flow velocity. 
    In panel~(c), the region rotating with the reference frame $\Omega_0$ is shown by white color. The iso-rotation
    contours are also overplotted.}
\end{figure*}
\begin{figure}[ht!]
  \epsscale{1.0}
  \plotone{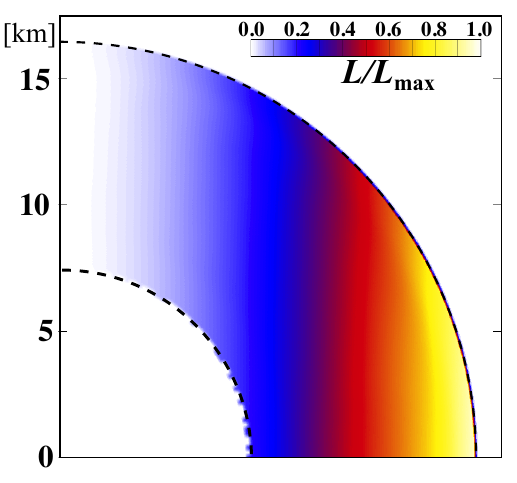}
  \caption{Meridional distribution of the normalized angular momentum, $\mathcal{L} \equiv \varpi^2\Omega$, in the northern hemisphere for {\it models}~ms (ms900p).
    The normalization unit is the maximum value of $\mathcal{L}$ which corresponds to that in the equatorial surface.}
\end{figure}
Unlike {\it models}~mf with the full-sphere CZ, the turbulent convective motion is confined within the spherical-shell in
{\it models}~ms as presented in figure~9(a), where the instantaneous snapshot for the profile of $v_r$ on a meridional cutting plane for
ms100p is demonstrated. The convection velocity becomes maximum at the upper part of the CZ as shown in figure~2(b), and
locally reaches $\mathcal{O}(10^9)\ {\rm cm/s}$. The symmetry of the system is broken by the spinning motion of the PNS, resulting
in an off-diagonal component of the Reynolds stress, which in turn gives rise to the mean flow and the large-scale thermodynamic
structure even in {\it models}~ms\footnote{When the large-scale magnetic field grows, the magnetic braking may carry the angular
momentum outside the CZ. However, as can be seen in Fig.6 or 10, there is no significant change in the angular velocity outside the CZ in our models.
In our model, with taking account of the existence of the convectively stable layer surrounding the CZ in the PNS, we put the damping region
outside the CZ. We control the flow velocity there to be almost the same as in the initial state (i.e., ${\bf v} = 0$), resulting in the redistribution
of the angular momentum only inside the CZ.}. 

Meridional distributions of (a) $\langle \langle v_\theta \rangle_\phi \rangle$,
(b) $\delta \epsilon \equiv \langle\langle \epsilon - \langle\langle{\epsilon}\rangle_s\rangle\rangle_\phi\rangle$
and (c) $\Omega \equiv \langle\langle v_\phi \rangle_\phi\rangle/r\sin\theta + \Omega_0$ for {\it models}~ms are shown in Figure~10.
The overplotted arrows in panel~(a) are meridional velocity vectors with arbitrary amplitudes. Panels~(a1)--(c1) are for ms100p,
(a2)--(c2) are for ms300p, and (a3)--(c3) are for ms900p, respectively. 

From the symmetry point of view, the large-scale flows seen in {\it models}~ms can be divided roughly into two types :
(i) multi--cell meridional circulations with a shellular differential rotation (ms100p), and (ii) a double-cell meridional
circulation (single-cell per hemisphere) with a cylindrical differential rotation (ms300p and ms900p). In ms100p with the lowest
spin rate, the multi-polar structure of the thermodynamic field is developed due to the multi-cell circulation flow. On the other hand,
the quadrupolar structure of the thermodynamic field, which is similar to that in mf60p and mf120p (see figs.6(b2)and(b3)), can be
observed in the models with relatively higher spin rates (ms300p and ms900p). 

In the case of thin spherical-shell convection, like that operating in the Sun, it is well-known that the multi-cell pattern of the
meridional flow is often observed in the faster rotation regime : the transition from the single-cell (per hemisphere) to multi-cell
profiles of the circulation occurs when the spin rate increases \citep[e.g.,][]{featherstone+15,mabuchi+15}. However, intriguingly, our PNS 
models with the spherical-shell convection appear to show the opposite behavior : single-cell (per hemisphere) pattern appeared in 
faster rotating models (ms300p and ms900p), while the multi-cell in slowly-rotating model (ms100p).

The single-cell circulation (per hemisphere) can be qualitatively understood by the so-called ``gyroscopic pumping'': when assuming
the quasi-steady state, the equation for the conservation of angular momentum can be derived from the zonal component of the 
mean-field equation of motion as
\begin{equation}
\overline{\rho {\bm v}_{\rm M} } \cdot \nabla \mathcal{L} = -\nabla \cdot \mathcal{\bm F}_{\rm RS} \;,
\end{equation}  
where the overbar denotes the time and longitudinal average, ${\bm v}_{\rm M}$ is the velocity of the meridional flow component,
$\mathcal{L} \equiv \varpi^2\Omega = \varpi\langle\langle v_\phi\rangle_\phi\rangle + \varpi^2\Omega_0$
is the specific angular momentum ($\varpi = r\sin\theta$ is the cylindrical radius), and
$\mathcal{\bm F}_{\rm RS}$ is the turbulent angular momentum flux by the convective Reynolds stress which is described as 
\begin{equation}
  \mathcal{\bm F}_{\rm RS} \equiv \overline{\rho}
  \varpi\left( \overline{ u_r' u_\phi'} {\bm e}_r + \overline{u_\theta' u_\phi'} {\bm e}_\theta \right) \;. 
\end{equation}
Note that the other non-axisymmetric component of the stress, such as the Maxwell stress and molecular viscosity,
are ignored because they are negligibly smaller than the Reynolds stress here.

Since the radial convection velocity is expected to be higher than the latitudinal and longitudinal velocities
in a relatively-slow rotation regime, the angular momentum is transported mainly in the radially inward direction,
that is $\mathcal{\bm F}_{\rm RS} \propto \overline{ u_r' u_\phi'} {\bm e}_r$ with $\overline{ u_r' u_\phi'} < 0$.
Then, the RHS of eq.(15) becomes positive (negative) in the upper (lower) CZ  because $\partial \mathcal{F}_{\rm RS} /\partial r > 0$
($\partial \mathcal{F}_{\rm RS} /\partial r < 0$) in the upper (lower) CZ when $\mathcal{F}_{\rm RS} = 0 $ at the top and bottom boundaries
\citep[e.g.,][]{miesch05,miesch+09,schrijver+10}. 

On the other hand, the $\mathcal{L}$ (specific angular momentum) has a cylindrical profile and decreases monotonically from equator to pole,
that is $\partial \mathcal{L}/\partial \theta < 0$ as shown in Figure~11 where the meridional distribution of the $\mathcal{L}$ in the northern
hemisphere of ms900p is demonstrated. Note that the similar profile of $\mathcal{L}$ can be seen in ms300p. To retain the zonal momentum balance
described by eq.(15), the meridional flow should be poleward (i.e., $\overline{v}_\theta <0$) in the upper CZ while it should be equatorward
(i.e., $\overline{v}_\theta >0$) in the lower CZ, providing the single-cell (per hemisphere) circulation flow seen in ms300p and ms900p.
Note that the meridional flow observed in mf60p and mf120p should be driven by the same mechanism, that is the gyroscopic pumping. 

The multi-cell pattern seen in ms100p (see fig.10(a1)) might be a consequence of the convective motion characterized by thick CZ, slow spin and relatively
low Reynolds number. According to \citet{chandra61} (\S~59), the spherical harmonic degree of the most unstable mode for the convective
instability depends on the thickness of the CZ and becomes lower with the increase of it. The higher the spherical harmonic degree of the mode,
the lower the growth rate becomes. At the limit of no rotation, the mode with $l=3$--$4$ seems to become the most unstable when the upper
$40$--$60$\% of the sphere is convective. In the system with small Reynolds and Rayleigh numbers, such as those expected in PNS, the growth
of the mode with larger spherical harmonic degree (i.e., $l \gtrsim 3$--$4$) and thus shorter wavelength tends to be suppressed. As a result,
the pattern of the linear unstable mode with lower spherical harmonic degree might be imprinted in the mean flow field like the circulation
flow seen in ms100p. 

The profile of the differential rotation in ms300p and ms900p is cylindrical (fig.10(c2) \& (c3)), that is in the Taylor-Proudman state like as
mf60p and mf120p, the flow being constrained to be quasi two-dimensional plane along the spin axis \citep[e.g.,][]{pedlosky82,kitchatinov+95}.
The lower the spin rate is, the weaker the rotational constraint on the flow becomes, resulting in the shellular rotation profile seen in ms100p (fig.10(c1)). 
Such a shellular rotation profile is also observed in \citet{brun+09} for the slowly-rotating red giant. As was described, in relatively-slowly
rotating systems, the angular momentum transport is mainly in the radially-inward direction, that is
$\mathcal{\bm F}_{\rm RS} \propto \overline{ u_r' u_\phi'} {\bm e}_r$ with $\overline{ u_r' u_\phi'} < 0$. The faster spinning motion in
the deeper CZ is thus a natural result of the turbulent transport process. 
\subsubsection{Magnetic Field : Dynamo Activities}
\begin{figure*}[ht!]
  \epsscale{0.8}
  \plotone{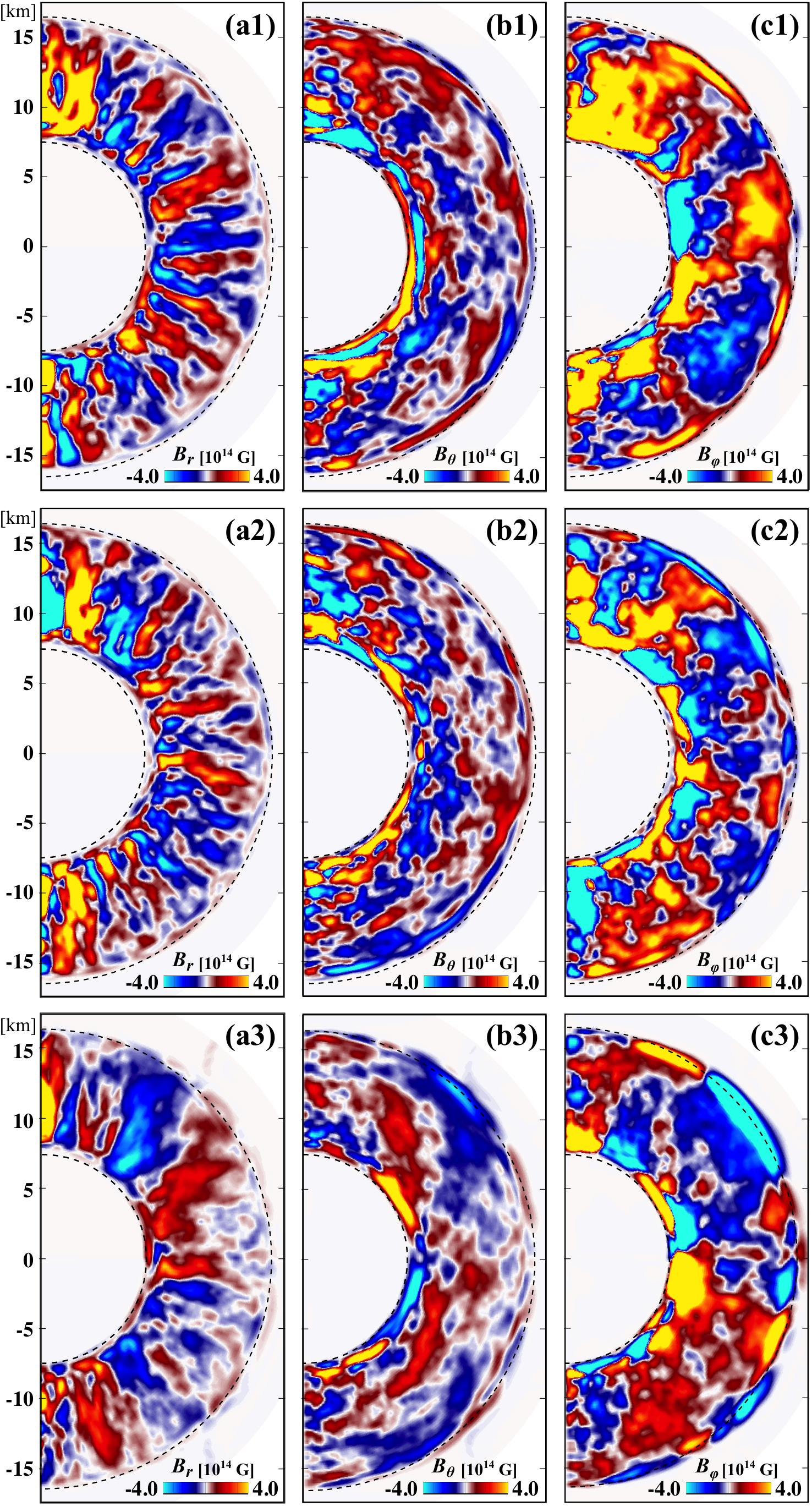}
  \caption{Meridional distributions of large-scale magnetic components, (a) $\langle\langle B_r \rangle_\phi\rangle$,
    (b) $\langle\langle B_\theta \rangle_\phi\rangle$, and (c) $\langle\langle B_\phi \rangle_\phi\rangle$ for {\it model}~ms.
    The top, middle and bottom panels are corresponding to {\it models}~ms100p, ms300p, and ms900p, respectively.
    The time average is taken over the duration $220 \le t \le 240$ ms.}
\end{figure*}
As shown in Figure~9(b), the spherial-shell convective motion amplifies turbulent magnetic fields of $\mathcal{O}(10^{16})$ G via small-scale dynamo process
\citep[e.g.,][]{cattaneo99,schekochihin+04}. The magnetic energy contained in the turbulent magnetic component is larger with the deeper depths.
When taking time and zonal average, we can find, even in {\it models}~ms, that the large-scale component of the magnetic field is developed in
these highly-disordered magnetic fields. 

In Figure~12, the meridional distributions of (a) $\langle\langle B_r \rangle_\phi\rangle$, (b) $\langle\langle B_\theta \rangle_\phi\rangle$ and (c)
$\langle\langle B_\phi \rangle_\phi\rangle$ for {\it models}~ms are shown. The top, middle and bottom panels are for {\it models}~ms100p,
ms300p, and ms900p, respectively. The time average is taken over the duration $220 \le t \le 240$ ms. The red (blue) tone denotes the positive (negative)
magnetic filed strength. 

As in {\it models}~mf, the mean component of the magnetic field has averagely a strength $\mathcal{O}(10^{14})$ G and locally exceeds $10^{15}$ G, which is comparable
to the field strength expected in ``magnetars''. When focusing on the geometry of the poloidal component of the magnetic field ($B_r$ and $B_\theta$), we can find it is
more complicated than that observed in the models with the full-sphere convection: while the higher multi-pole structure becomes dominant in the models in relatively-slow
rotation regime (ms100p and ms300p), the quadrupolar dominance is prominent in the fastest spinning model (ms900p). Additionally to the poloidal component, a large-scale
toroidal magnetic component is also built up in all models. Commonly, it is roughly anti-symmetric with respect to the equator and seems to have a strength which is
a bit ``stronger'' than the poloidal magnetic component. The dependence of the larges-scale magnetic field on the spin rate and the mechanism for the large-scale dynamo
observed in our simulation models are discussed in the following section in detail.   
\section{Discussion}
\subsection{Rotational Dependence of Large-scale Magnetic Field}
\begin{figure}[ht!]
  \epsscale{1.17}
  \plotone{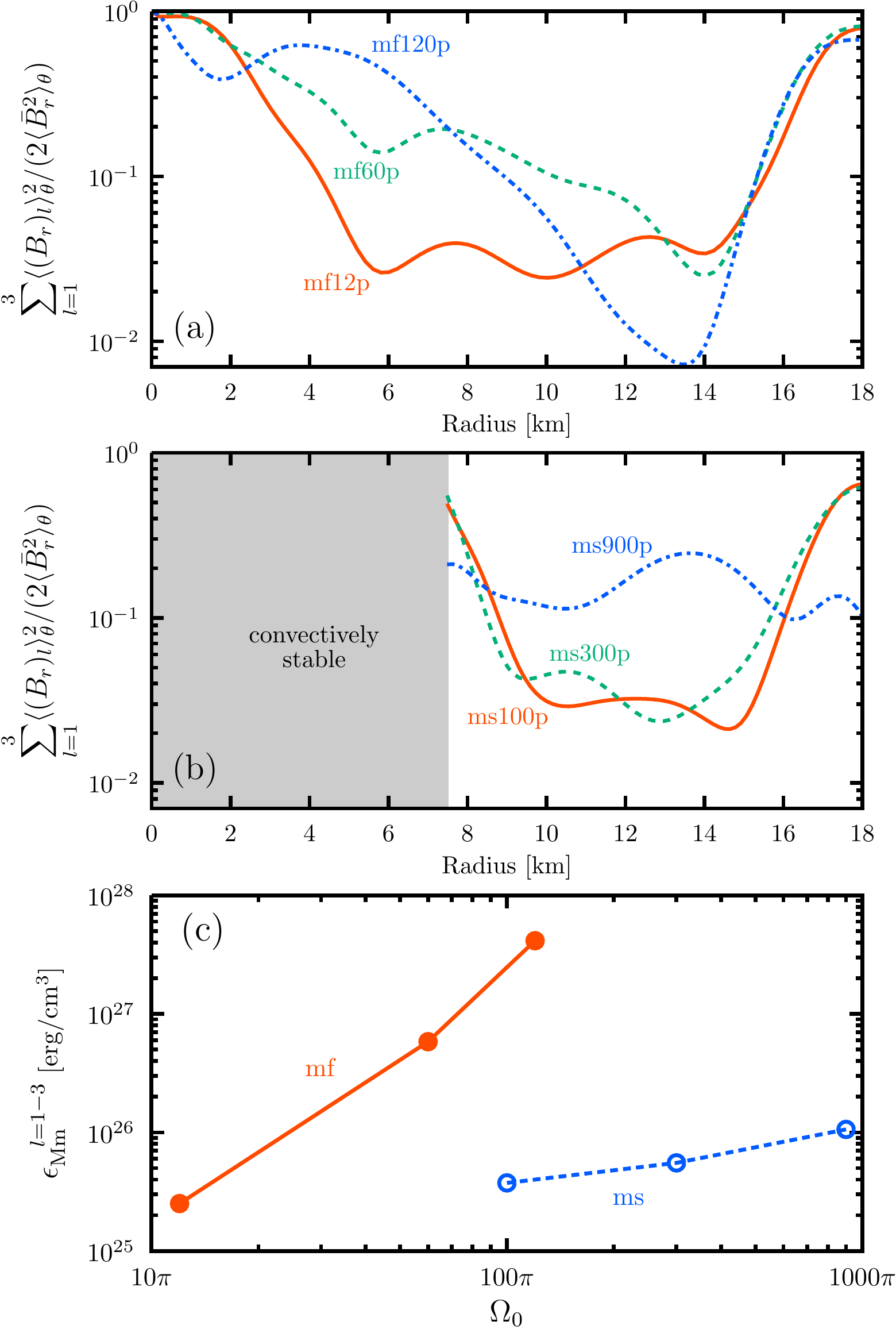}
  \caption{Radial distributions of the dominance of the large-scale field for (a) {\it models}~mf and (b) {\it models}~ms. The different line types
    denote models with different spin rates. Panel~(c) shows the dependence of the mean magnetic energy of the large-scale magnetic filed on the spin rate
    for {\it model}~mf (red-solid) and {\it model}~ms (blue-dashed). }
\end{figure}
As presented in Figs.~8 and 12, the mean component of the magnetic field is spontaneously organized in all the PNS models we examined in this paper.
Here we discuss the dependence of the structure and strength of the mean magnetic component on the spin rate of the PNS systematically. 

To gain quantitative insights into it, the latitudinal moments of the axisymmetric field, $\bar{\bm B}$, is analyzed where the overbar, used to simplify the notation,
denotes the time and longitudinal average, i.e., $\bar{B}_r = \langle \langle B_r \rangle_\phi\rangle$. We focus on the radial component $\bar{B}_r$ since it
purely reflects the poloidal field, while $\bar{B}_\theta$ and $\bar{B_\phi}$ are a mixture of the toroidal and poloidal fields. From the Perseval's equation,
a relation 
\begin{equation}
  \langle \bar{B}_r^2 \rangle_\theta = \frac{1}{2}\sum_{l=1}\left( \bar{B}_r \right)_l^2 \;,
\end{equation}
where
\begin{equation}
\left(\bar{B}_r \right)_l = \int_{-1}^{1} \bar{B}_r P_l^*(\cos\theta){\rm d}\cos\theta \;, 
\end{equation}
holds \citep[see][for details]{masada+13}. Here $P_l^*$ are normalized Legendre polynomials. In the following, we focus on three larger scale
modes $l=1,2$, and $3$ (dipole, quadrupole, and octapole) as an indicator of the efficiency of the large-scale dynamo. 

Shown in Figure~13 is the radial distribution of the ratio of the magnetic energy stored in the largest-scale components ($l=1$--$3$)
to the total magnetic energy of the axisymmetric field, that is $ \sum_{l=1}^3\left(\bar{B}_r \right)_l/2\langle\bar{B}_r^2\rangle_\theta $
for (a) {\it models}~mf and (b) {\it models}~ms. The different line types denote the models with different spin rates. The gray shaded
region in panel~(b) corresponds to the convectively-stable region of {\it models}~ms. Note that the higher the ratio, the more the
large-scale field becomes dominant.

For {\it models}~mf (panel~(a)), the dominance of the large-scale component in the magnetic energy is more pronounced in the deeper CZ and decreases with the
radius, suggesting that the dynamo effect is stronger in the deeper depth. Interestingly, even in mf12p with the smallest spin rate (red-solid line), the
central part of the PNS ($r \lesssim 5$ km) is occupied by the large-scale component, which is consistent with the result of Figure~8. As the spin rate increases,
the region where the large-scale magnetic component becomes dominant extends into the outer part of the CZ, implying that the dynamo effect becomes stronger in
the higher spin rate of the PNS. 

Comparing {\it models}~mf and ms, the dominance of the large-scale magnetic component seems to be weaker in the {\it models}~ms.
It is a common feature of all the models that the strong large-scale component is developed near the upper and lower boundaries, but in the mid-part of
the CZ, only the model ms900p shows a strong large-scale magnetic component (blue dash-dotted line in panel~(b)). This implies that the large-scale dynamo
can be excited in the mid-part of the CZ only when the PNS's spin is fast enough. 

Figure~13(c) shows the dependence, on the spin rate, of the mean magnetic energy stored in the large-scale component defined by
\begin{equation}
  \epsilon_{{\rm Mm},l=1-3} = \sum_{l=1}^3\left[ \int_{r_{\rm min}}^{r_{\rm max}}\left(\bar{B}_r \right)_l^2 {\rm d}r \bigg/ \int_{r_{\rm min}}^{r_{\rm max}} {\rm d}r \right]\;,
\end{equation}
where $r_{\rm min}$ and $r_{\rm max}$ is the pseudo- upper and lower boundaries of the CZ. We choose $r_{\rm min} = 0$ km and $r_{\rm max} = 17$ km for {\it models}~mf
and $r_{\rm min} = 7.5$ km and $r_{\rm max} = 17.5$ km for {\it models}~ms.

Commonly in both models, we can find a clear tendency for the mean magnetic energy stored in the large-scale component to increase with the spin rate.
However, there exists a difference in the strength of the dependency on the spin rate. In the regime we studied in this paper, the slope of the spin-dependence is
steeper in {\it models}~mf than in {\it models}~ms. It would be necessary to investigate the dependence of the PNS dynamo on the spin rate in a wider
parameter range to understand the difference in the trends. 

\subsection{Dynamo Mechanism}
\begin{figure}[ht!]
  \epsscale{1.15}
  \plotone{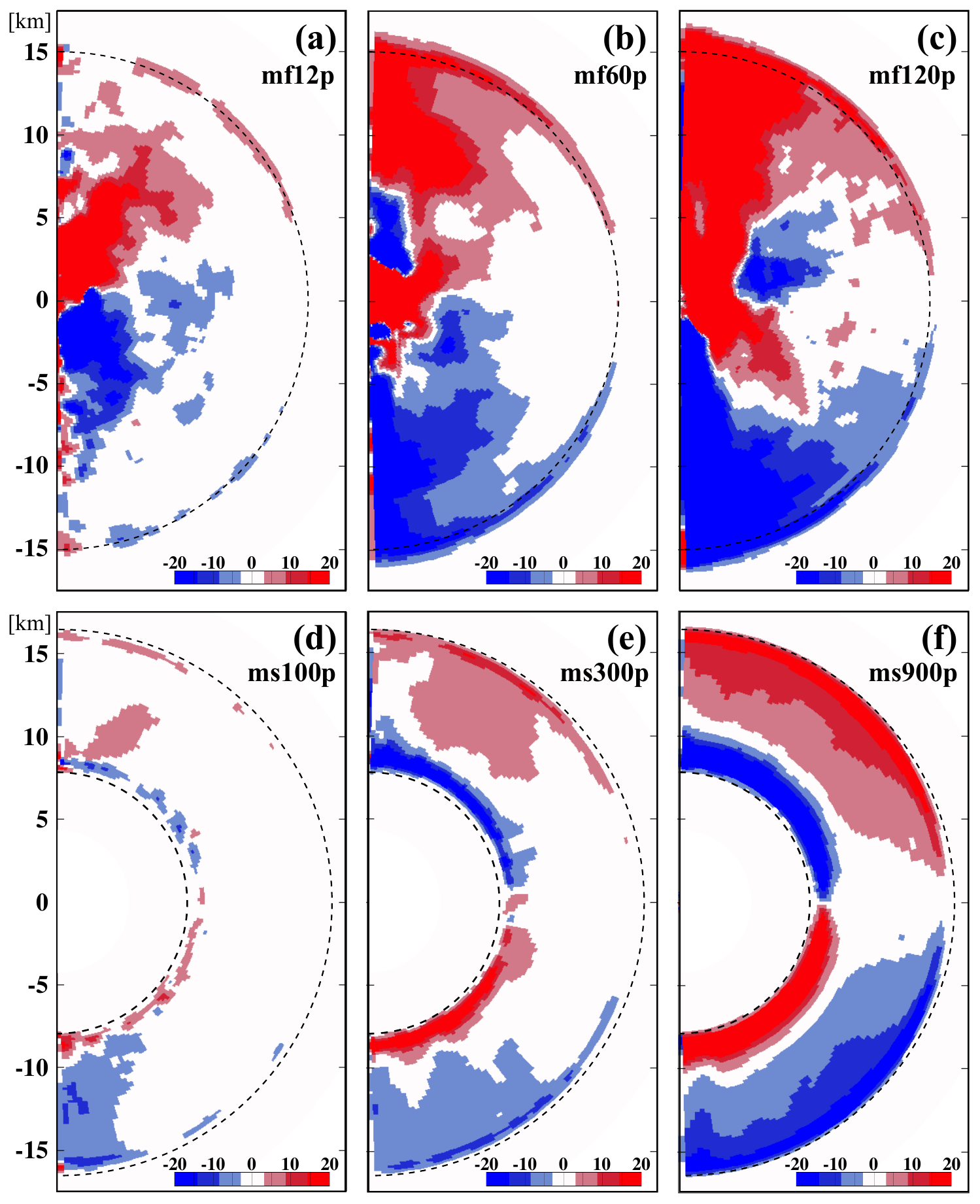}
  \caption{Meridional distributions of $\mathcal{C}_\alpha = \alpha H_\rho/\eta_t$ for {\it models}~mf [(a)--(c)] and {\it models}~ms [(d)--(f)].}
\end{figure}
It is well known that the rotating convection system spontaneously generates a mean kinetic helicity with a north-south anti-symmetry, i.e., in the
case of the eastward spinning motion like our PNS model, bulk negative helicity in north and positive in south, because of the Coriolis force
acting on the convection flow \citep[e.g.,][]{miesch05,miesch+09}. From preceding studies on stellar and solar dynamos \citep[e.g.,][for reviews]{charbonneau14,brun+17,charbonneau20},
we expect that the turbulent electro-motive fore (EMF) would be the key for generating the large-scale magnetic component even in our PNS models
\citep[e.g.,][]{racine+11,masada+14}. 

Although it is difficult to evaluate quantitatively the role of the turbulent EMF in the complicated PNS models, we can
appraise it at least qualitatively based on the mean-field dynamo (MFD) theory. Under the first-order smoothing approximation
\citep[e.g.,][]{brandenburg+05,masada+14}, the kinetic helicity would be closely linked to the turbulent $\alpha$-effect, which is a
key ingredient for inducting the large-scale magnetic field in the MFD framework \citep[e.g.,][]{moffatt78,krause+80}, as
\begin{equation}
\alpha \equiv -\tau_{\rm cor} \langle\langle \mbox{\boldmath $v$}'\cdot \mbox{\boldmath $\omega$}' \rangle_\phi\rangle /3 \;,
\end{equation}
where $\mbox{\boldmath $v$}' \equiv \mbox{\boldmath $v$} - \langle\langle \mbox{\boldmath $v$}\rangle_\phi\rangle$ is the turbulent
velocity, $\tau_{\rm cor}$ is the correlation time and $\mbox{\boldmath $\omega$}' \equiv \nabla \times \mbox{\boldmath $v$}'$ is the
turbulent vorticity. Additionally to it, the turbulent magnetic diffusivity, which parameterizes the turbulent
transport of the magnetic energy through advection and reconnection and thus controls the destruction process of the large-scale magnetic field,
is linked to the turbulent velocity as 
\begin{equation}
\eta_t \equiv \tau_{\rm cor} \langle\langle \mbox{\boldmath $v$}'^2 \rangle_\phi\rangle/3 \;. 
\end{equation}
Essentially, these parameterizations for $\alpha$ and $\eta_t$ are valid only in isotropic turbulence \citep[e.g.,][]{schrijver+10}. However,
even in an-isotropic turbulence, these have been shown to be useful in studying the ability of a system to excite the large-scale dynamo
\citep[e.g.,][]{racine+11,masada+14}.

In eqs.(20) and (21), the correlation time can be evaluated as the convective turn-over time, that is 
$\tau_{\rm cor} = H_\rho/\langle v_r^2 \rangle^{1/2}$, where $H_\rho$ is the density scale-height. Since $\mbox{\boldmath $v$}'$, $\mbox{\boldmath $\omega$}'$
and $H_\rho$ can be extracted from the simulation data directly, we can depict meridional distributions of the turbulent $\alpha$ and turbulent magnetic
diffusivity for each model. In the following, we discuss the hidden connection between the large-scale magnetic component observed in our PNS models
and the turbulent EMF with using a non-dimensional parameter $\mathcal{C}_\alpha$ defined by
\begin{equation}
\mathcal{C}_\alpha = \alpha H_\rho /\eta_t \;, 
\end{equation}
which is equivalent to the so-called ``Dynamo number'' when the $H_\rho$ is chosen as the typical spatial scale on which the large-scale dynamo works. 
Since the $\alpha$ and $\eta_t$ describes the induction and destruction effects of the large-scale magnetic component respectively, the amplitude of
$\mathcal{C}_\alpha$ becomes a measure of the efficiency of the large-scale dynamo. 

Shown in Figure~14 is the meridional distribution of $\mathcal{C}_\alpha$ which is derived directly from the simulation data obtained in each model.
Panels~(a)--(c) correspond to those in {\it models}~mf and panels~(d)--(f) are for {\it models}~ms. Note that the color scale
is the same for all panels. The darker the color, the stronger the relative induction effect becomes. 

Since the kinetic helicity has an anti-symmetric profile with respect to the equator (bulk negative in north and positive in south),
the profile of $\mathcal{C}_\alpha$ also shows the quasi-anti-symmetry between hemispheres. With comparing Figs~13 and 14, it can be found
that there exists a remarkable overlap between the region with the strong large-scale magnetic component, and the region with the large
$|\mathcal{C}_\alpha|$ in both models. In addition to that, the region with the large $|\mathcal{C}_\alpha|$ extends into the outer part
of the CZ like as that observed in the large-scale magnetic component. Comparing two models, we can see that the amplitude of $C_\alpha$
is larger in {\it models}~mf than in {\it models}~ms, suggesting that the higher efficiency of the large-scale dynamo in {\it models}~mf.
This is consistent with the stronger large-scale magnetic component in {\it models}~mf seen in Fig.13(c). Overall these results suggest
that the turbulent EMF plays an important role in the large-scale dynamo in our PNS models.

Then, a natural question arises ``{\it why is the efficiency of the large-scale dynamo higher in models~mf though
the spin rate of {\it models}~mf is smaller than that of {\it models}~ms ? }''. The key for it would be
the typical size and velocity of the convective motion. The turbulent $\alpha$-effect can be brought to the form
\begin{equation}
\alpha = -\frac{1}{3}\tau_{\rm cor}^2\mbox{\boldmath $v$}'^2 \left[{\bm \Omega}\cdot\nabla\ln(\overline{\rho\mbox{\boldmath $v$}'})\right] \;,
\end{equation}
\citep[e.g.,][]{steenbeck+69,charbonneau13} if we assume that the inhomogeneity arises from the stratification and the symmetry breaking from the Coriolis force, and the
lifetime of turbulent eddies is evaluated by the their turnover time. With this description, the dynamo number $\mathcal{C}_\alpha$
can be rewritten as
\begin{eqnarray}
  \mathcal{C}_\alpha & = & \alpha H_\rho /\eta_t \nonumber \\
  & \simeq & \Omega H_\rho \cos\theta /v_{\rm rms} \propto Ro^{-1} \;, 
\end{eqnarray}
with a rough estimation, $\nabla\ln(\overline{\rho\mbox{\boldmath $v$}'}) \simeq 1/H_\rho$. When ignoring the latitudinal dependence,
we can see that the dynamo number, which is a measure of the efficiency of the large-scale dynamo, is a function of the spin rate,
the scale-height, and the velocity of the turbulent convective motion. 

On the whole, {\it models}~mf are assumed to have lower spin rates than in {\it models}~ms. On the other hand, as shown in figs.~2 and 5, 
{\it models}~mf has a lower convection velocity and a larger size of the convective eddies especially in the central part of the PNS.
These results indicate that the larger $\mathcal{C}_\alpha$ realized in {\it models}~mf is a consequence of the smaller convection velocity
and larger size of the convective eddies (especially in the deeper CZ), which compensate for its smaller spin rate, resulting in the
environment more suitable for the large-scale dynamo. It can be said that, physically, the deeper the CZ extends, the larger the size
of the convection eddies, and thus the rotationally-constrained convection is more easily achieved, resulting in a region more suitable 
for the large-scale dynamo. 
\section{Summary}
In this paper, we constructed a ``PNS in a box'' simulation model with solving the compressible MHD with a nuclear EOS and a simplified leptonic
transport to study properties of MHD convection and dynamo in PNSs. As a demonstration of our newly-developed model, we applied it to two types
of the internal structure of the PNS : fully-convective state and spherical-shell convection state. Our main findings are summarized as follows.

1. The large-scale flows developed in {\it models}~mf are divided into two types : (i) a single-cell meridional circulation with 
a north-south anti-symmetric differential rotation (mf12p), and (ii) a double-cell meridional circulation (single-cell per hemisphere) with
a cylindrical differential rotation (mf60p and mf120p). While the dipole dominance in $\delta \epsilon$ is accompanied with the flow
pattern~(i), the quadrupolar dominance in $\delta \epsilon$ is developed in conjunction with the pattern~(ii).

2. The large-scale flows developed in {\it models}~ms are also divided into two types : (i) multi–cell meridional circulations with a
shellular differential rotation (ms100p), and (ii) a double-cell meridional circulation (single-cell per hemisphere) with a cylindrical
differential rotation (ms300p and ms900p). While the multi-polar structure of $\delta \epsilon$ is accompanied with the flow pattern~(i),
the quadrupolar structure of $\delta \epsilon$ is developed in the models with relatively higher spin rates.

3. With taking account of the angular momentum transport due to the turbulent Reynolds stress caused by rotating convective motions, the circulation
pattern of the formed meridional flow can be qualitatively understood by the so-called ``gyroscopic pumping''. On the other hand, the profile of the
differential rotation is determined to maintain the thermal wind balance of the system. 

4. The magnetic field amplified by the PNS convection shows a complicated structure mixed with the turbulent and large-scale components.
It should be emphasized that, in all the PNS models we studied here, the large-scale component of the magnetic field is spontaneously organized.
For {\it models}~mf, the mean magnetic component exhibits a dipole dominance, i.e., the large-scale poloidal component, rooted deep in the central
part of the PNS, shows a strong dipole symmetry. In contrast to that, for {\it models}~ms, while the higher multi-pole structure becomes dominant
in the models in a relatively-slow rotation regime, the quadrupolar dominance is prominent in the fastest spinning model.

5. Although there exists a clear tendency for the mean magnetic energy stored in the large-scale component to increase with the spin rate in both models,
the slope of the spin-dependence is steeper in {\it models}~mf than in {\it models}~ms. Additionally, it is intriguing that, as an overall trend,
{\it models}~mf has a stronger large-scale magnetic component than that in {\it models}~ms. 

6. There exists a remarkable overlap between the region with the strong large-scale magnetic component, and the region with the large $|\mathcal{C}_\alpha|$
in both models, where $\mathcal{C}_\alpha$ is equivalent to the so-called ``dynamo number'' and a measure of the efficiency of the large-scale dynamo.
Comparing two models, the amplitude of $\mathcal{C}_\alpha$ is larger in {\it models}~mf than in {\it models}~ms, suggesting that the higher efficiency
of the large-scale dynamo in {\it models}~mf. Since the deeper the CZ extends, the larger the size of the convection eddies, the rotationally-constrained
convection seems to be more easily achieved in {\it models}~mf. As a result, the full-sphere convection state becomes more suitable for the large-scale dynamo.

Although the convective dynamo is believed conventionally to work only in rapidly rotating PNSs (e.g., TD93), a clear dynamo activity
can be found even in a slowly-rotating PNS with $P_{\rm rot} \simeq 170$ ms (mf12p). This would be essentially due to the setup of
{\it models}~mf in which the CZ extends to the deeper part of the PNS. It is well-known that the width of the CZ in the PNS changes
depending not only on the physical properties of the progenitor star \citep[e.g.,][]{nagakura+19} but also on the evolution phase of
the PNS. As demonstrated in the 2D hydrodynamic simulation of the deleptonization of the PNS by \citet{keil+96}, the CZ in the PNS
enlarges to the deeper part with the progress of the neutrino cooling, finally encompassing the whole star within $\sim 1$ s after
bounce, and can continue for at least as long as the deleptonization takes place (see also, \citet{roberts+12}).

Since most of the existing studies for the PNS dynamo supposes the early evolutionary stage at which only the outer part of the PNS is
convective, the scale-height, thus the size of the convective eddies, is relatively small there, resulting in the high $\mathcal{C}_\alpha$
in the case with the slow or moderate rotation (e.g., TD93). On the other hand, at the later evolutionary stage, the PNS is expected to have
a deeper CZ with a larger size of the convective eddies in the deeper part like the fully-convective models we studied here. In such a situation,
the Coriolis force dominates over the inertia force around the PNS core, and thus the large-scale magnetic component might be efficiently
amplified there by the turbulent $\alpha$-effect against the turbulent magnetic diffusion, i.e., the region with $\mathcal{C}_\alpha \gg 1$ might 
be more easily developed. Overall our results imply that the PNS dynamo may become more efficient in later phases of its evolutionary stage. 
We note that the importance of the deep core convection for the large-scale dynamo has already been pointed out in the study for the
origin of the magnetic field in the fully-convective M-type dwarfs \citep[][]{yadav+16,kapyla21}.

Recently, \citet{beniamini+19} studied the formation rate of Galactic magnetars directly from observations and estimated that a fraction
of $0.4^{+0.6}_{-0.28}$ of NSs are born as magnetars with magnetic fields of $B \gtrsim 3\times 10^{13}$ G. This finding is a challenge to standard
scenarios for the magnetar formation, as these senarios require more or less extreme conditions, such as pre-collapse rapid rotation and/or strong magnetic
fields. As a physical mechanism to explain a possibly high fraction of the magnetars in the NSs, \citet{soker20} proposed the stochastic omega-effect
(S$\Omega$ effect) and claimed that the rapid rotation is not necessarily required for the magnetar formation. Our studies in this paper also suggest
that the strong magnetic fields expected in the magnetars can be organized in newly-born NSs even at relatively slow rotations, which may prompt a
reconsideration of the existing scenarios for the magnetar formation.

To build up the concrete view on the role of the PNS convection in the context of the origin of
NS's magnetic fields, we should study the MHD convection in the PNS in the wider parameter range with varying the depth of the CZ, diffusivities,
the rotation rate, and the initial strength of the magnetic field, respectively. These are beyond the scope of this paper, but will be a target of
our future work.
\begin{acknowledgments}
  We acknowledge the anonymous referee for constructive comments. This work has been supported by MEXT/JSPS KAKENHI Grant numbers
  JP17H01130, JP17K14306, JP18H01212, JP18K03700, JP17H06357, JP17H06364, JP18H04444, JP21K03612;
the Central Research Institute of Stellar Explosive Phenomena (REISEP) at Fukuoka University and the associated projects (Nos.\ 171042,177103)
and JICFuS as a priority issue to be tackled by using Post `K' Computer.
Numerical computations were carried out on Cray XC50 at Center for Computational Astrophysics, National Astronomical Observatory of Japan.
\end{acknowledgments}

\end{document}